\documentclass[10pt,aps,onecolumn,floats,floatfix,amssymb,prd,superscriptaddress,nofootinbib,, preprintnumbers]{revtex4-2}

\usepackage{graphicx}
\usepackage{amsmath,amssymb}
\usepackage{amsfonts}
\usepackage{xspace} 
\usepackage[usenames]{color}
\usepackage{dcolumn}
\usepackage{bm}
\usepackage{mathrsfs}
\usepackage[colorlinks=true]{hyperref}
\usepackage[all]{hypcap} 
\usepackage[utf8]{inputenc} 
\usepackage{multirow}
\usepackage{etoolbox}
\usepackage{tikz}
\usepackage[normalem]{ulem}

\newcommand{\dd}{\mathrm{d}}

\begin{document}

\title{Proca theory of four-dimensional regularized Gauss-Bonnet gravity\\and black holes with primary hair}

\author{Christos Charmousis}
\email{christos.charmousis@ijclab.in2p3.fr}
\affiliation{Université Paris-Saclay, CNRS/IN2P3, IJCLab, 91405 Orsay, France}
\affiliation{Theoretical Physics Department, CERN, 1211 Geneva 23, Switzerland}

\author{Pedro G. S. Fernandes}
\email{fernandes@thphys.uni-heidelberg.de}
\affiliation{Institut f\"ur Theoretische Physik, Universit\"at Heidelberg, Philosophenweg 12, 69120 Heidelberg, Germany}

\author{Mokhtar Hassaine}
\email{hassaine@inst-mat.utalca.cl}
\affiliation{Instituto de Matemáticas, Universidad de Talca, Casilla 747, Talca, Chile}
\preprint{CERN-TH-2025-080}
\begin{abstract}
We introduce a novel, well-defined four-dimensional regularized Gauss-Bonnet theory of gravity by applying a dimensional regularization procedure. The resulting theory is a vector-tensor theory within the generalized Proca class. We then consider the static spherically symmetric solutions of this theory and find black hole solutions that acquire primary hair. Notably, one of the integration constants associated with the Proca field is not manifest in the original metric, but under a disformal transformation of the seed solution, it emerges as a second, independent primary hair. This additional hair acts as an effective cosmological constant in the disformed geometry, even in the absence of a bare cosmological constant term. We further generalize these black hole solutions to include electromagnetic charges and effects related to the scalar-tensor counterparts of the regularized Gauss-Bonnet theory. We discuss the implications of our findings to observations.
\end{abstract}

\maketitle

\section{Introduction}
Lovelock theories of gravity \cite{Lovelock:1971yv,Padmanabhan:2013xyr} are prominent higher-curvature extensions of General Relativity (GR). They generalize the Einstein-Hilbert action by including higher-order curvature scalars while retaining second-order equations of motion and avoiding additional degrees of freedom. The $D$-dimensional Lovelock action consists of a linear combination of the first $\lceil D/2 \rceil$ Lovelock invariants, as higher-order terms either vanish or become topological. In four dimensions, the theory reduces to GR with a cosmological constant, whereas in five or more dimensions, the leading correction arises from the Gauss-Bonnet (GB) term.  

The GB term is notable for several reasons beyond being the first non-trivial Lovelock correction to GR. In four dimensions, it reduces to the Euler density and thus becomes a topological invariant, leaving the dynamics unaffected. However, when non-minimally coupled to a scalar field, it yields non-trivial contributions to the field equations. Crucially, it is the only higher-curvature term that can couple to a scalar field without introducing higher-derivative equations of motion, and is thus part of the broad Horndeski class \cite{Horndeski:1974wa,Kobayashi:2019hrl}. Moreover, GB contributions routinely occur in the low-energy limit of quantum theories of gravity, such as string theory \cite{Ferrara:1996hh, Antoniadis:1997eg, Zwiebach:1985uq, Nepomechie:1985us, Callan:1986jb, Candelas:1985en, Gross:1986mw}.

Recent years have witnessed significant interest in the GB term and its applications, and in particular in the development of well-defined four-dimensional theories that reproduce solutions of higher-dimensional Einstein-Gauss-Bonnet gravity\footnote{A generalised Kaluza-Klein reduction of Lovelock gravity \cite{Charmousis:2012dw} results in four-dimensional black hole solutions but these do not have the usual four-dimensional Newtonian fall-off due to their higher-dimensional origin.}. Such theories, known as 4D Einstein-Gauss-Bonnet (4DEGB) theories, were long considered impossible due to the topological nature of the GB invariant in four dimensions. However, this perspective shifted following the work of Glavan and Lin \cite{glavanEinsteinGaussBonnetGravity4dimensional2020}, who demonstrated that through a singular rescaling of the GB coupling constant within a dimensional regularization scheme, one can obtain a non-trivial 4DEGB theory whose solutions mirror those of higher-dimensional Lovelock gravity.

The proposal in Ref. \cite{glavanEinsteinGaussBonnetGravity4dimensional2020} was ultimately shown to be ill-defined \cite{gursesThereNovelEinsteinGaussBonnet2020}. Nevertheless, it inspired subsequent works \cite{luHorndeskiGravityRightarrow42020,kobayashiEffectiveScalartensorDescription2020,fernandesDerivationRegularizedField2020,hennigarTakingLimitGaussBonnet2020} to develop a consistent dimensional regularization procedure. This led to well-defined scalar-tensor theories within the Horndeski class, featuring a scalar field with generalized conformal symmetry \cite{Fernandes:2021dsb}. Remarkably, these theories often admit the same solutions as those found in the original, now invalidated approach. For a comprehensive review of 4DEGB gravity, see Ref. \cite{fernandes4DEinsteinGaussBonnetTheory2022} and its references.

In electrovacuum, a stationary black hole is fully characterized by its mass $M$, angular momentum $J$, and electric charge $Q_e$ \cite{PhysRev.174.1559,PhysRevLett.34.905,Mazur:1982db,Chrusciel:2012jk}. Indeed, any two such black holes with identical values of $M$, $J$, and $Q_e$ are described by the same exact Kerr–Newman solution. Owing to the absence of any additional externally observable physical quantities, black holes are colloquially said to have \textit{no hair}.
However, in alternative theories of gravity, black holes can possess hair. Black hole hair is typically classified as either \emph{primary} or \emph{secondary}. In the secondary case, there are no new global charges -- independent from $M$, $J$, and $Q_e$ -- associated with a Gauss law. Nevertheless, the spacetime geometry deviates from the Kerr–Newman form, even though it remains entirely determined by these three charges. This scenario arises in the vast majority of modified gravity theories, including the 4DEGB models discussed in Refs. \cite{luHorndeskiGravityRightarrow42020,kobayashiEffectiveScalartensorDescription2020,fernandesDerivationRegularizedField2020,hennigarTakingLimitGaussBonnet2020,Fernandes:2021dsb}. A much less common situation involves primary hair, where the black hole is characterized not only by $M$, $J$, and $Q_e$, but also by additional independent global charges. Theories in which black holes possess primary hair are particularly compelling, as the spacetime geometry is modified not only by new coupling constants of the theory but also by independent integration constants that can vary across different physical settings. The feature of primary hair opens the possibility, for example, for substantial deviations from the Kerr paradigm in the geometry of supermassive black holes -- something generally precluded in most theories with secondary hair, due to stringent constraints on the length scale associated with the new couplings \cite{Eichhorn:2023iab}. Consequently, black holes with primary hair open the possibility that deviations from the Kerr geometry can arise across a wide range of black hole masses within the same theoretical framework. This has far-reaching implications for various observational efforts, from gravitational wave detections \cite{LIGOScientific:2016aoc,LIGOScientific:2017vwq,LIGOScientific:2018mvr,amaro2017laser,Barausse:2020rsu} to very-long-baseline interferometry experiments \cite{Ayzenberg:2023hfw,EventHorizonTelescope:2019ggy,EventHorizonTelescope:2020qrl,EventHorizonTelescope:2021dqv,EventHorizonTelescope:2022xqj}. We highlight that black hole solutions featuring primary hair remain exceptionally rare, with only a few known instances in the literature. Notable examples include the solutions of Refs. \cite{Herdeiro:2014goa,Herdeiro:2016tmi}, where gravity is minimally coupled to a complex bosonic field, allowing for rotating black holes with primary hair. More recent developments include solutions found in the context of beyond-Horndeski theories \cite{Bakopoulos:2023fmv,Baake:2023zsq,Bakopoulos:2023sdm,Charmousis:2025xug}, as well as in Refs. \cite{Gervalle:2024yxj,Gervalle:2025awa} (see also references therein), which explore scenarios involving ``standard model" hair.

In this paper, we carry out a dimensional regularization of the GB term to construct a novel 4DEGB theory, which differs from those in Refs. \cite{luHorndeskiGravityRightarrow42020,kobayashiEffectiveScalartensorDescription2020,fernandesDerivationRegularizedField2020,hennigarTakingLimitGaussBonnet2020}. This is achieved by introducing a vector field in the framework of Weyl geometry, rather than a scalar field. This idea was first suggested in the conclusions of Ref. \cite{Fernandes:2021dsb}, but had not been realized until now. The resulting theory is a vector–tensor theory within the generalized Proca class \cite{Heisenberg:2014rta}, and is therefore free from Ostrogradsky instabilities. We investigate the black hole solutions of this theory and find that, while they closely resemble those of the higher-dimensional Einstein–GB theory, they intriguingly feature primary, rather than secondary hair, rendering them particularly compelling and worthy of further exploration.

The paper is organized as follows. In Sec. \ref{sec:derivetheory}, we review previous dimensional regularizations of the GB term to four dimensions, which lead to scalar–tensor theories. We then introduce the framework of Weyl geometry and show how it naturally gives rise to vector–tensor theories with enhanced conformal properties. Using this framework, we construct a well-defined four-dimensional vector–tensor (Proca) GB theory. In Sec. \ref{sec:bhs}, we analyze the black hole solutions of this theory and show that they exhibit primary hair. This is due to two free integration constants associated to the Proca field. The first integration constant directly modifies the geometry, while the second is only present in the Proca field configuration. In order to understand the physical implications of the latter, we consider a disformal transformation of the black hole solution. The disformed solution is again a black hole, but where the second integration constant acts as another primary hair. Interestingly, this primary hair now provides an effective cosmological constant even in the absence of a bare cosmological constant. Next, we extend the black hole solutions to include electromagnetic charges and consider the effects of combining vector–tensor and scalar–tensor regularized theories. We present our conclusions in Sec. \ref{sec:conclusions}, along with a discussion of possible extensions of our work.

\section{Proca Four-Dimensional Einstein-Gauss-Bonnet theory}
\label{sec:derivetheory}

We start this section by reviewing the regularization procedure done in Refs. \cite{fernandesDerivationRegularizedField2020,hennigarTakingLimitGaussBonnet2020} to get a 4DEGB theory, and their connection to the results of Ref. \cite{Fernandes:2021dsb}. This will serve as a guide to construct the new 4DEGB theory of this article. The regularization procedure of Refs. \cite{fernandesDerivationRegularizedField2020,hennigarTakingLimitGaussBonnet2020} consists on using two conformally related metrics, $\hat{g}_{\mu \nu} = e^{2\phi} g_{\mu \nu}$, and the following limit
\begin{equation}
    \sqrt{-g}\mathcal{L}_\mathcal{G}^{\rm ST} = \lim_{d\to 4} \frac{\sqrt{-\hat{g}} \hat{\mathcal{G}} - \sqrt{-g} \mathcal{G}}{d-4},
    \label{eq:limit_ST}
\end{equation}
where
\begin{equation}
    \mathcal{G} = R^2 - 4 R_{\mu \nu} R^{\mu \nu} + R_{\mu \nu \alpha \beta} R^{\mu \nu \alpha \beta},
\end{equation}
is the GB invariant, and ``ST" stands for ``scalar-tensor". The limit can be shown to be well-defined \cite{Colleaux:2020wfv}, and results in the following scalar-tensor theory with second-order equations of motion \cite{fernandesDerivationRegularizedField2020,hennigarTakingLimitGaussBonnet2020}
\begin{equation}
    \mathcal{L}_\mathcal{G}^{\rm ST} = \phi \mathcal{G} - 4 G^{\mu \nu} \nabla_\mu \phi \nabla_\nu \phi - 4 \Box \phi (\partial \phi)^2 - 2 (\partial \phi)^4.
\end{equation}
This Lagrangian, when supplemented with the Einstein-Hilbert term results in the 4DEGB theory
\begin{equation}
    S = \int \dd^4x \sqrt{-g} \left( R - \beta \mathcal{L}_\mathcal{G}^{\rm ST} \right),
    \label{eq:4DEGB_ST}
\end{equation}
where $\beta$ is the GB coupling constant with dimensions of length squared. This theory also gives compact neutron star solutions with interesting phenomenology \cite{Charmousis:2021npl,Saavedra:2024fzy} as well as explicit and well-defined wormhole spacetimes \cite{Bakopoulos:2021liw,Babichev:2022awg}, among other interesting features \cite{Fernandes:2021ysi}. It also reproduces many of the solutions of the ill-defined procedure of Ref. \cite{glavanEinsteinGaussBonnetGravity4dimensional2020}, for example in black holes and cosmology, and they are analogous to those of the higher-dimensional theory. 

As shown in Ref. \cite{Fernandes:2021dsb}, the good integrability of theory \eqref{eq:4DEGB_ST} can be traced to enhanced properties of the theory under the conformal transformation
\begin{equation}
    g_{\mu \nu} \to e^{2\sigma} g_{\mu \nu}, \quad \phi \to \phi - \sigma,
    \label{eq:conformal_ST}
\end{equation}
where $\sigma \equiv \sigma(x)$ is a conformal factor. This can be seen already at the level of the limit in Eq. \eqref{eq:limit_ST}, by noting that any scalar constructed from the metric $\hat{g}_{\mu \nu}$ and its derivatives is invariant under the transformation in Eq. \eqref{eq:conformal_ST}.

In Ref. \cite{Fernandes:2021dsb}, it was suggested that vector–tensor theories with similar properties could offer interesting possibilities. This is the direction we pursue in the present work. In particular, Eq. \eqref{eq:conformal_ST} generalizes to the vector-tensor case as
\begin{equation}
    g_{\mu \nu} \to e^{2\sigma} g_{\mu \nu}, \quad W_\mu \to W_\mu - \partial_\mu \sigma,
    \label{eq:conformal_VT}
\end{equation}
where $W_\mu$ is a new vector field. Theories with enhanced properties with respect to the transformation \eqref{eq:conformal_VT} can be realized naturally within the framework of Weyl geometry \cite{Weyl:1918ib} involving the use of a Weyl rather than a Levi-Civita connection. We shall follow Refs. \cite{jimenezExtendedGaussBonnetGravities2014,bahamondeExactFiveDimensional2025,barceloWeylRelativityNovel2017} closely when discussing Weyl geometry.

In GR, the parallel transport of vector directions is non-integrable, whereas the transport of vector lengths is integrable. In the framework of Weyl geometry the latter is rendered non-integrable as well, by introducing a new torsionless connection $\tilde{\Gamma}^{\lambda}\,_{\mu \nu}$ with an associated covariant derivative $\tilde{\nabla}_\mu$ such that
\begin{equation}
    \tilde{\nabla}_{\lambda}g_{\mu \nu}=-2g_{\mu\nu}W_{\lambda}\,,
\end{equation}
where $W_\mu$ is a vector field.
The components of the Weyl connection on a coordinate basis are
\begin{equation}
    \tilde{\Gamma}^{\lambda}\,_{\mu \nu}=\Gamma^{\lambda}\,_{\mu \nu} + \delta^{\lambda }{}_{\mu } W_{\nu } + \delta^{\lambda }{}_{\nu } W_{\mu } - g_{\mu \nu } W^{\lambda },
\label{eq:WeylConnection}
\end{equation}
where $\Gamma^{\lambda}\,_{\mu \nu}$ is the Levi-Civita connection, and the Levi-Civita covariant derivative satisfies $\nabla_{\lambda}g_{\mu \nu} = 0$. We are now in a position to observe that the Weyl connection \eqref{eq:WeylConnection} remains invariant under the conformal transformation given in Eq. \eqref{eq:conformal_VT}, and consequently, all curvature tensors constructed from the Weyl connection transform covariantly under this transformation. For example, the Riemann tensor of the Weyl connection
\begin{equation}
    \tilde{R}^{\lambda}\,_{\rho \mu \nu}=\partial_{\mu}\tilde{\Gamma}^{\lambda}\,_{\rho \nu}-\partial_{\nu}\tilde{\Gamma}^{\lambda}\,_{\rho \mu}+\tilde{\Gamma}^{\lambda}\,_{\sigma \mu}\tilde{\Gamma}^{\sigma}\,_{\rho \nu}-\tilde{\Gamma}^{\lambda}\,_{\sigma \nu}\tilde{\Gamma}^{\sigma}\,_{\rho \mu}\,,
\end{equation}
is invariant under Eq. \eqref{eq:conformal_VT}.

Of particular interest to us, is the GB invariant. Ref. \cite{jimenezExtendedGaussBonnetGravities2014} discusses in great detail how to construct the equivalent of the GB invariant for the Weyl connection (see also Ref. \cite{bahamondeExactFiveDimensional2025}). Upon simplifying the expression in Ref. \cite{jimenezExtendedGaussBonnetGravities2014}, the GB term for the Weyl connection can be brought to the form\footnote{The GB term for the Weyl connection coincides with the one obtained from a conformal transformation of the metric only, upon carefully identifying the derivative of the conformal factor with the vector field.}
\begin{equation}
    \begin{aligned}
    \tilde{\mathcal{G}}=& \mathcal{G} + (d-3)\nabla_\mu J^\mu + (d-3)(d-4) \mathfrak{L}\,,
    \end{aligned}
    \label{eq:GBconf0}
\end{equation}
where
\begin{equation}
    \begin{aligned}
        &J^\mu = 8 G^{\mu \nu} W_\nu + 4(d-2)\left[W^\mu(W^2 + \nabla_\nu W^\nu) - W_\nu \nabla^\nu W^\mu\right],\\&
        \mathfrak{L} = 4 G^{\mu \nu}W_\mu W_\nu + (d-2)\left( 4 W^2 \nabla_\mu W^\mu + (d-1)W^4 \right),
    \end{aligned}
    \label{eq:GBconf}
\end{equation}
and where we have used the shorthand notation $W^2=W_\mu W^\mu$, $W^4 = (W^2)^2$.

Equipped with the expression for the GB invariant of the Weyl connection, we are ready to perform the dimensional regularization of the GB term
\begin{equation}
    \mathcal{L}_\mathcal{G}^{\rm VT} = \lim_{d\to 4} \frac{\tilde{\mathcal{G}}-\mathcal{G}}{d-4},
\end{equation}
which, upon discarding total derivatives, evaluates to
\begin{equation}
    \boxed{\mathcal{L}_\mathcal{G}^{\rm VT} = 4 G^{\mu \nu}W_\mu W_\nu + 8 W^2 \nabla_\mu W^\mu + 6W^4.}
\end{equation}
Similarly to the scalar-tensor case, we supplement the theory with the Einstein-Hilbert term, and therefore the 4DEGB vector-tensor theory is given by\footnote{We adopt this sign for the coupling constant $\alpha$ to align our results with those of the scalar–tensor theory, as will become evident in the following section.}
\begin{equation}
    S = \int \dd^4x \sqrt{-g} \left(R   -\alpha \mathcal{L}_\mathcal{G}^{\rm VT} \right),
    \label{eq:theory}
\end{equation}
where $\alpha$ is now the GB coupling constant, with dimensions of length squared.
Note that the theory is a well-defined four-dimensional theory belonging to the generalized Proca class \cite{Heisenberg:2017mzp,Heisenberg:2014rta} with
\begin{equation}
    G_2 = -24\alpha X^2, \quad G_3 =- 16 \alpha X, \quad G_4 = 1 -4\alpha X,
    \label{procahorn}
\end{equation}
where $X = -W_\mu W^\mu/2$, and therefore its equations of motion are at most second-order. They are presented in Appendix \ref{appA}. The part of the action containing the vector field is invariant under (global) conformal transformations, while the Einstein-Hilbert term is not. As explained in Ref. \cite{Babichev:2024krm} this leads to a current whose divergence is sourced by the Ricci scalar
\begin{equation}
    R = \frac{\alpha}{2} \nabla_\mu J^\mu,
\end{equation}
where $J^\mu$ is defined in Eq. \eqref{eq:GBconf}.

\section{Black holes with primary hair}
\label{sec:bhs}

In this section we study the asymptotically flat, static and spherically symmetric black hole solutions of the theory \eqref{eq:theory}. We consider the line-element given by
\begin{equation}
    \dd s^2 = -N(r)^2f(r) \dd t^2 + \frac{\dd r^2}{f(r)} + r^2\left( \dd \theta^2 + \sin^2\theta \dd \varphi^2 \right),
    \label{eq:metric}
\end{equation}
and the vector field
\begin{equation}
    W_\mu \dd x^\mu = w_0(r) \dd t + w_1(r) \dd r.
    \label{eq:ansatzVF}
\end{equation}
We use an effective Lagrangian approach where we insert our ansatz in the action, and vary with respect to $N$, $f$, $w_0$ and $w_1$ (see Appendix \ref{appA}). A trivial solution is the Schwarzschild metric together with a vanishing vector. However, we seek non-trivial solutions. To that end, we note that the equations of motion in this situation have the symmetry (see Appendix \ref{appA})
\begin{equation}
    w_0^2 \to w_0^2 + \kappa\left(2rw_1fN+r^2\kappa+2fN\right), \qquad w_1 \to w_1 + \kappa\frac{r}{Nf},
    \label{eq:symmetry}
\end{equation}
where $\kappa$ is an arbitrary constant. Since Eq. \eqref{eq:symmetry} is a continuous symmetry of the static and spherically symmetric system, there is an associated first integral/conserved quantity
\begin{equation}
    \frac{\dd}{\dd r} \left[ r\left( 1 - (1+r w_1)^2f + \frac{r^2 w_0^2}{f N^2} \right) \right] \equiv \frac{\dd}{\dd r} \mathcal{Q} = 0.
    \label{eq:consQuant}
\end{equation}
From Eq. \eqref{eq:consQuant} we can derive a profile for $w_0$ in terms of $r$, the other functions, and the conserved quantity. Using this profile of $w_0$, the equations of motion $\delta S/\delta w_0$ and $\delta S/\delta w_1$ become equivalent. Then, a specific combination of $\delta S/\delta w_0$ with the equation $\delta S/\delta f$ imposes $N' = 0$, from which we get $N=1$ without loss of generality. At this point, the equations $\delta S/\delta w_0$, $\delta S/\delta w_1$ and $\delta S/\delta f$ are all equivalent to one another, and can be solved to give $w_1$ in terms of $r$, $f$, the conserved quantity and a new integration constant $c$. We are left only with the equation $\delta S/\delta N$, that can be integrated analytically, where yet another integration constant, $M$, appears. The final result can be expressed as\footnote{Similarly to other GB-related solutions, there are two branches of solutions, only one of which is asymptotically flat, which is the one we consider. The other branch can be obtained by flipping the sign before the square-root in Eq. \eqref{sol1}.}

\begin{equation}
    f(r) = 1 - \frac{2(M-Q)}{r} + \frac{r^2}{2\alpha} \left( 1-\sqrt{1+\frac{8\alpha Q}{r^3} } \right),
    \label{sol1}
\end{equation}
\begin{equation}
    w_0^2=g^2+2 c f, \qquad w_1=\frac{g}{f},
\end{equation}
where,
\begin{equation}
    g(r)=\frac{1-f}{2r}-\frac{M-Q}{r^2}+cr=\frac{2 Q}{r^2}\left(1+\sqrt{1+\frac{8\alpha Q}{r^3}}\right)^{-1}+cr,
    \label{procac}
\end{equation}
and where we have defined the conserved quantity $\mathcal{Q}$ in terms of $Q$, which we take to be our parameter measuring hairyness, as $\mathcal{Q} = 2(M-Q)$.
Expanding $f$ in \eqref{sol1} for large $r$ it is easy to see that $M$ is the ADM mass of the black hole, whereas $Q$ and $c$ are related to the Proca field. Note that the Proca function $g(r)$ in Eq. \eqref{procac} does not depend on $M$ and is uniquely characterised by the charges $c$ and $Q$. The integration constant $c$ does not modify the geometry, but on the other hand, $Q$ acts as primary hair since it is not fixed in terms of the mass $M$ and does modify the geometry. Interestingly, on-shell we have that the norm of the Proca field is constant, $X=c$. When $c=0$, both components of the vector field behave asymptotically as $Q/r^2 + \mathcal{O}(r^{-3})$.

For this solution, the symmetry in Eq. \eqref{eq:symmetry} becomes $g\to g + \kappa r$, and $c\to c + \kappa$.
This symmetry explains why $c$ does not modify the geometry, since it can be set to zero by choosing $\kappa=-c$. The transformation also changes $X \to c + \kappa$.

When $Q$ is set to zero, we recover a stealth Schwarzschild black hole. Indeed, note that the Proca vector is not trivial due to the presence of $c$. When we set $Q=M$ we have the black hole geometry of the scalar-tensor versions of 4DEGB \cite{luHorndeskiGravityRightarrow42020,kobayashiEffectiveScalartensorDescription2020,fernandesDerivationRegularizedField2020,hennigarTakingLimitGaussBonnet2020,Fernandes:2021dsb}.  In general, as we will now see, the solution \eqref{sol1} describes an asymptotically flat black hole.

Indeed when $\alpha<0$ we have a black hole which is quite similar to a Schwarzschild black hole apart that its horizon size is increased as we increase the magnitude of negative $\alpha$. Furthermore, for $\alpha<0$, we have a singularity at finite $r=r_{\rm min}$ where $r_{\rm min}=(-8 \alpha Q)^{1/3}$. On the other hand if $\alpha>0$ our horizon is on the contrary more compact, than in the GR case, $Q=0$. As we increase the charge $Q$ and we  approach $Q=M$ we start having a triple and then a double horizon. For large enough $Q=Q_{\rm ext}$, our inner and event horizon coincide and we have an extremal black hole. Beyond this point $Q>Q_{\rm ext}$ we have a naked singularity. This latter behavior is quite similar to the 4DEGB case (compare with the analysis in \cite{Charmousis:2021npl}). Lastly, if we have $M=0$ and non trivial Proca charge $Q\neq 0$ two different possibilities arise: either we have a black hole horizon at $r=r_h$, for $Q$ relatively large as compared to $\alpha<0$, so as $r_h>r_{\rm min}$. Otherwise we have a naked singularity as one has for a Reissner-Nordstrom geometry in GR.

\begin{figure}[!h]
    \includegraphics[width=0.5\textwidth]{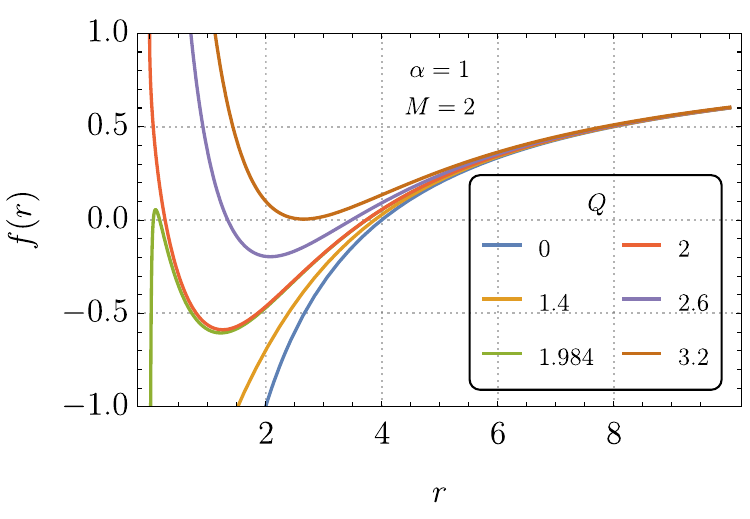}
    \caption{Sketch of the black hole metric function $f$ given in \eqref{sol1} for $\alpha=1$, $M=2$ and different values for the primary hair parameter $Q=\{0,0.7M,0.992M, M, 1.3M, 1.6M\}$.}
    \label{fig:1}
\end{figure}

Interestingly, since $Q$ is not fixed in terms of the mass (primary hair), even when the GB coupling is small $ \rvert\alpha\rvert /M^2\ll 1$, we can have non-trivial important modification of the geometry. For example, consider the small coupling regime but where the primary hair is large enough such that $\rvert\alpha\rvert Q^2/M^4 = \eta \sim \mathcal{O}(1)$. Then, to first order in $M/Q$, we have
\begin{equation}
    f(r) = 1 - \frac{2M}{r} + \frac{4M^4 \eta}{r^4} \mathrm{sgn}(\alpha) + \mathcal{O}(M/Q).
\end{equation}
This result is impactful, for example, for supermassive black hole observations, since it provides a mechanism where the geometry of supermassive black holes can differ substantially from the Schwarzschild metric, providing a counterexample to the prevailing expectation that deviations from the Schwarzschild geometry increase with horizon curvature \cite{Eichhorn:2023iab}.

\subsection{A black hole with a cosmological constant as primary hair}

In the previous section we observed that the second integration constant, $c$, does not modify the spacetime metric. To understand the physical implications of this Proca constant we now consider a disformal transformation of the solution in Eq. \eqref{sol1}.

Disformal transformations for Proca theories and resulting constraints from gravitational wave observations have been studied in \cite{Domenech:2018vqj}.
A disformal transformation 
\begin{equation}
\label{disformal0}
\bar g_{\mu \nu}=g_{\mu \nu}+D\; W_\mu W_\nu\,,   
\end{equation}
of our initial theory \eqref{procahorn}
involving some constant $D$, will transform our seed vector-tensor solution \eqref{sol1} to an image solution in a vector-tensor theory which is again a Proca theory with transformation rules for  $\bar G_2, \bar G_3$ and $\bar G_4$  found in \cite{Domenech:2018vqj}. The constancy of $D$ guarantees \cite{Domenech:2018vqj} that a disformal transformation is an internal map within $G_2, G_3, G_4$.  Unlike scalar tensor theories, where the disformed solution, for $X$ constant, is just a coordinate transformation of the seed metric (see for example \cite{Babichev:2017lmw, BenAchour:2020wiw}), here we have a radically different spacetime. What is of particular interest to us, is that the integration constant $c$ now explicitly appears in the metric.  Indeed, let us without further a do display the spacetime solution where by barred quantities we will be denoting the image Proca spacetime. We emphasize that the solution belongs to the space of solutions for the disformed Proca theory which can be found directly in \cite{Domenech:2018vqj}.

Performing the disformal transformation \eqref{disformal0}, the disformed spherically symmetric spacetime reads
\begin{equation}
    \dd s^2 = -\bar f(r) \dd \tau^2 + \frac{\dd r^2}{\bar f(r)} + r^2\left( \dd \theta^2 + \sin^2\theta \dd \varphi^2 \right),
    \label{eq:metricdis}
\end{equation}
where our rescaled time is defined as, $\dd t= \frac{1}{1-2D X} \dd \tau + \frac{D w_0 w_1}{f-D w_0^2} \dd r$, making the spacetime metric diagonal. 
The disformed metric function reads
\begin{equation}
\bar{f}=\frac{f-D w_0^2}{1-2D X}
\label{fxd}
\end{equation}
while the Proca vector remains unchanged. 
We can also expand the latter relation to get
\begin{equation}
 \bar f=f - \frac{D}{2(1-2c D)r \alpha} \left( Q + 2c^2 r^3 \alpha - \frac{2Q (1-4\alpha c)}{1+\sqrt{1+\frac{8Q\alpha}{r^3}}} \right)
\end{equation}
and therefore our spacetime is now asymptotically de Sitter rather than flat as long as $c$ is non zero.  

To make this clear we can set the second charge $Q=0$, and we have,
\begin{equation}
\label{de}
 \bar f=1 -\frac{2M}{r}-\frac{Dc^2}{(1-2c D)}r^2.   
\end{equation}
We see that now $c$ acts as a primary hair and plays the role of an effective cosmological constant of magnitude
\begin{equation}
\Lambda_{\rm eff}=\frac{Dc^2}{3(1-2c D)}. 
\label{Lambda}
\end{equation}
The stringent constraints provided by multi-messenger gravitational wave data \cite{LIGOScientific:2017zic, LIGOScientific:2017vwq} will relate the values of $c$ to the coupling constant $\alpha$ and the constant $D$. In this sense $c$ will be reduced to secondary hair.
Indeed, the speed of tensor modes for our disformed theory is given by \cite{Domenech:2018vqj,Dong:2023xyb}
\begin{equation}
    \bar c_T^2=\frac{c_T^2}{1-2D X}=
    \frac{1-4 \alpha X}{(1+4 \alpha X)(1-2 D X)}
\end{equation}
Since our solution changes $\Lambda_{\rm eff}$ \eqref{Lambda} at cosmological distances, in accord with \cite{LIGOScientific:2017vwq, LIGOScientific:2017zic} we set $\bar c_T^2 \sim 1$ which gives
\begin{equation}
    c\sim \frac{1}{D}-\frac{1}{4\alpha}, \qquad \Lambda_{\rm eff} \sim \frac{4c^2 \alpha}{3-12 c \alpha}.
\end{equation}
The equations above tell us that, for any effective cosmological constant $\Lambda_{\rm eff}$ given by Eq. \eqref{Lambda} there exists a ``physical" disformal frame defined by $D$ where $\bar c_T^2\sim 1$ in agreement with observational data~\cite{LIGOScientific:2017vwq, LIGOScientific:2017zic}.

\subsection{Combining the scalar-tensor and the vector-tensor theories}
It is possible to combine the regularized vector-tensor theory in Eq. \eqref{eq:theory} with the scalar-tensor theory in Eq. \eqref{eq:4DEGB_ST}, and still get analytic black hole solutions with primary hair. The theory we consider now is\footnote{We note that the scalar-tensor part could be further extended as in Ref. \cite{Fernandes:2021dsb} or \cite{Charmousis:2021npl} in a very similar fashion.}
\begin{equation}
    S = \int \dd^4x \sqrt{-g} \left(R   -\alpha \mathcal{L}_\mathcal{G}^{\rm VT} - \beta \mathcal{L}_\mathcal{G}^{\rm ST} \right),
    \label{eq:theorySTVT}
\end{equation}
where $\beta$ is a coupling constant with the same dimensions as $\alpha$.
Using the profiles for the primary hair solution of the vector-tensor theory, together with the scalar field profile \cite{fernandesDerivationRegularizedField2020,hennigarTakingLimitGaussBonnet2020}
\begin{equation}
    \phi^{\prime} = \frac{1-\sqrt{f}}{r\sqrt{f}},
\end{equation}
we obtain the following solution
\begin{equation}
    f(r) = 1-\frac{2 \alpha(M-Q)}{r(\alpha+\beta)}+\frac{r^2}{2(\alpha+\beta)} \Bigg( 1-\sqrt{1+\frac{8\alpha Q}{r^3} +\frac{8\beta M}{r^3}- \frac{16 (M-Q)^2 \alpha \beta}{r^6} } \Bigg),
\end{equation}
if $\beta \neq -\alpha$, or
\begin{equation}
    f(r) = 1 - \frac{2M r^2 - \frac{4(M-Q)^2 \alpha}{r}}{r^3 - 4 (M-Q) \alpha}=\frac{r^3}{r^3 - 4 (M-Q) \alpha}\Bigg[1-\frac{2M}{r}+\frac{4(M-Q)^2\alpha}{r^4}-\frac{4(M-Q)\alpha}{r^3}\Bigg],
\end{equation}
when $\beta=-\alpha$. In the particular case when $Q=M$ and $\beta=-\alpha$,  the metric reduces to the Schwarzschild solution, and thus the configuration corresponds to a stealth black hole. On the other hand, when $\beta=0$ we recover the solution in Eq. \eqref{sol1}. These solutions generically describe black holes, and are similar to those found in Ref. \cite{bahamondeExactFiveDimensional2025} in the context of higher dimensions. The solutions have a curvature singularity at $r=0$, or at the location where the quantity inside the square-root vanishes, although this singularity is generically shielded by a horizon.

\subsection{Charged black holes with primary hair}
For completeness, we also present the charged generalizations of the black holes discussed in previous sections. To this end, we consider the action
\begin{equation}
    S = \int \dd^4x \sqrt{-g} \left(R - F_{\mu \nu}F^{\mu \nu}  - \alpha \mathcal{L}_\mathcal{G}^{\rm VT} - \beta \mathcal{L}_\mathcal{G}^{\rm ST} \right),
    \label{eq:theorySTVTcharged}
\end{equation}
where $F_{\mu \nu} = \partial_\mu A_\nu -\partial_\nu A_\mu$, where $A_\mu$ is a $U(1)$ gauge field. We consider again the line-element \eqref{eq:metric}, and
\begin{equation}
    A_\mu \dd x^\mu = a_0(r) \dd t - Q_m \cos \theta \dd \varphi,
\end{equation}
where $Q_m$ is the magnetic charge. The derivation of the charged solution follows as in previous sections. The Maxwell equations impose
\begin{equation}
    a_0 = \frac{Q_e}{r} + \Phi_e,
\end{equation}
where $Q_e$ is the electric charge and $\Phi_e$ the electrostatic potential. Using all the profiles derived before, we obtain the solution
\begin{equation}
    f(r) = 1-\frac{2(M-Q) \alpha}{r(\alpha+\beta)}+\frac{r^2}{2(\alpha+\beta)} \Bigg( 1-\sqrt{1+\frac{8\alpha Q}{r^3}+ \frac{8\beta M}{r^3}- \frac{16 (M-Q)^2 \alpha \beta}{r^6} - \frac{4 (Q_e^2+Q_m^2)}{r^4}\left( \alpha + \beta \right) } \Bigg),
\end{equation}
if $\beta \neq -\alpha$, or
\begin{equation}
    \begin{aligned}
        f(r) &= 1 - \frac{2M r^2 - 4(M-Q)^2 \frac{\alpha}{r} - (Q_e^2 + Q_m^2)r}{r^3 - 4 (M-Q) \alpha}\\&=\frac{r^3}{r^3 - 4 (M-Q) \alpha}\Bigg[1-\frac{2M}{r}+\frac{(Q_e^2 + Q_m^2)}{r^2}+\frac{4\alpha (M-Q)^2}{r^4}-\frac{4\alpha(M-Q)}{r^3}\Bigg],
    \end{aligned}
\end{equation}
when $\beta=-\alpha$. It is interesting to note that for $Q=M$, we end up with a stealth defined on the top the Reissner-Nordstrom solution. 

%%%%%%%%%%%%%%%%%%%%%%%%%
\section{Conclusions}
\label{sec:conclusions}
%%%%%%%%%%%%%%%%%%%%%%%%%%

In this work, we have derived a novel, well-defined four-dimensional Einstein–Gauss–Bonnet theory using a dimensional regularization procedure involving a vector field. In contrast to previously established 4DEGB theories \cite{luHorndeskiGravityRightarrow42020,kobayashiEffectiveScalartensorDescription2020,fernandesDerivationRegularizedField2020,hennigarTakingLimitGaussBonnet2020,fernandes4DEinsteinGaussBonnetTheory2022,Fernandes:2021dsb}, which are scalar–tensor in nature, the theory presented here is a vector–tensor theory\footnote{See Refs. \cite{Barton:2021wfj,Barton:2022rkj} for works that considered vector-tensor Gauss-Bonnet theories in other contexts, and Refs. \cite{Heisenberg:2017xda, Heisenberg:2017hwb} for works that studied hairy black holes in the generalized Proca class.}. Our construction is motivated by the findings of Ref. \cite{Fernandes:2021dsb}, which demonstrated that all known scalar–tensor 4DEGB theories possess a scalar field with improved conformal properties. This symmetry leads to improved integrability and has enabled the discovery of exact analytic solutions, such as black holes \cite{fernandes4DEinsteinGaussBonnetTheory2022,Babichev:2022awg,Babichev:2023dhs} or other compact objects \cite{Charmousis:2021npl, Bakopoulos:2021liw}. Vector-tensor theories with improved conformal properties can be constructed naturally in the context of Weyl geometry, which we considered here to this end.

Following Refs. \cite{jimenezExtendedGaussBonnetGravities2014,bahamondeExactFiveDimensional2025}, we have shown that the GB invariant constructed from the Weyl connection can be expressed as the GB term computed with the Levi-Civita connection, plus a total derivative and a term that vanishes in four dimensions (see Eq. \eqref{eq:GBconf0}), as it is proportional to a factor of $(d-4)$. Employing this decomposition, along with a regularization procedure inspired by Refs. \cite{fernandesDerivationRegularizedField2020,hennigarTakingLimitGaussBonnet2020}, we demonstrated that a well-defined four-dimensional vector–tensor 4DEGB theory can be constructed (see Eq. \eqref{eq:theory}). The resulting theory belongs to the well-established generalized Proca class \cite{Heisenberg:2014rta}, and is thus free from Ostrogradsky instabilities.

We have investigated the black hole solutions of this vector–tensor 4DEGB theory and found that, in contrast to its scalar–tensor counterparts, the theory admits black holes with primary hair, representing a maximal violation of the no-hair conjectures. This primary hair emerges as one (of the two)  of integration constants and crucially, is not determined by the other global charges, such as the black hole mass, or by the coupling constants of the theory. For completeness, we also obtained generalizations of these black hole solutions that incorporate electromagnetic charge, as well as configurations arising when the vector–tensor and scalar–tensor 4DEGB theories are combined. 

In order to understand the physical implications of the second free integration constant we considered a disformal transformation of our black hole solution. The resulting theory is again a Proca theory involving disformed $\bar G_2, \bar G_3, \bar G_4$ functionals as was found in \cite{Domenech:2018vqj}. Interestingly we found that the integration constant which was absent in the seed metric now appears as primary hair playing the role of a free cosmological constant. This may revive  interesting self-tuning scenarios as those put forward by Fab 4 theories \cite{Charmousis:2011bf, Charmousis:2011ea, Appleby:2012rx} and subsequent efforts \cite{Appleby:2018yci, Babichev:2016kdt} that could be interesting to study in the near future. The interesting new feature in this self-tuning scenario is that the self tuning parameter is not a priori fixed by the coupling constants of the theory. As such it can be fixed by gravitational wave constraints \cite{LIGOScientific:2017vwq, LIGOScientific:2017zic}
with respect to the disformal frame for any effective cosmological constant.

In this direction we can try to understand indicatively the cosmologies of the theory at hand \eqref{eq:theory}. For this, one can employ a flat Friedmann-Lemaître-Robertson-Walker, consider matter content in the form of a perfect fluid, and consider a homogenous vector field which does not break spatial isotropy, given by $W_\mu \dd x^\mu = -w_0(t) \dd t$. Under these conditions, it is possible to find that, assuming a non-trivial solution, the vector field equation of motion imposes $w_0 = H$, where $H$ is the Hubble parameter. When substituted in the Einstein equations we obtain the generalized Friedmann equation $H^2 + \alpha H^4 = \frac{8\pi}{3}\rho$, which has exactly the same form as the ones from the higher-dimensional GB theory \cite{Deruelle:1989fj}, and coincide also with the ones obtained in the scalar-tensor theory \eqref{eq:4DEGB_ST} \cite{Fernandes:2021dsb,fernandes4DEinsteinGaussBonnetTheory2022}. Moreover, these Friedmann equations have been derived in other contexts, such as in holographic cosmology \cite{Apostolopoulos:2008ru,Bilic:2015uol}, by considering a generalized uncertainty principle \cite{Lidsey:2009xz}, or by considering a quantum corrected entropy-area relation of the apparent horizon of a FLRW universe \cite{Cai:2008ys}. For an in-depth study of the phenomenology obtained from these generalized Friedmann equations, the reader is referred to Ref. \cite{fernandes4DEinsteinGaussBonnetTheory2022}.

As a direction for future research, it would be interesting to explore the black hole solutions derived here in greater depth. Since the presence of primary hair can significantly affect the geometry -- even for supermassive black holes -- it becomes important to investigate the image features associated with these spacetimes, particularly in light of observations by the Event Horizon Telescope collaboration. Additionally, a detailed analysis of the quasi-normal modes of these black holes is warranted. We note that the quasi-normal modes of other black holes with primary hair \cite{Bakopoulos:2023fmv} were recently studied in Ref. \cite{Charmousis:2025xug}. The thermodynamics of these black holes also present an important and interesting avenue for future research.

%%%%%%%%%%%%%%%%%%%%%%%%%%
\section*{Acknowledments}
%%%%%%%%%%%%%%%%%%%%%%%%%%
PF is funded by the Deutsche Forschungsgemeinschaft (DFG, German Research Foundation) under Germany’s Excellence Strategy EXC 2181/1 - 390900948 (the Heidelberg STRUCTURES Excellence Cluster). CC acknowledges partial 
support of ANR grants StronG (ANR-22-CE31-0015-01) and COSQUA  (ANR-20-CE47-0001). 
We are very happy to thank Aimeric Colléaux for numerous and enlightening discussions on Lovelock theories and regularization procedures, and Eloy Ayón-Beato and Eugeny Babichev for useful comments on the manuscript. PF would like to thank the theory department at CERN for their kind hospitality during the initial stages of this work. CC happily acknowledges discussions with Karim Noui on Proca theories. MH gratefully acknowledges the University of Paris-Saclay for its kind hospitality during the development of this project.

%%%%%%%%%%%%%%%%%%%%%%%%%%%%%%%%%%%%%%%%
\appendix
\section{Field equations and a symmetry in spherical symmetry}\label{appA}
%%%%%%%%%%%%%%%%%%%%%%%%%%%%%%%%%%%%%%%%
The field equations of the action \eqref{eq:theory} are given by 
\begin{eqnarray*}
&&G_{\mu\nu}=-\alpha \mathcal{T}_{\mu\nu},\nonumber\\
\nonumber\\
&& G^{\mu}_{\space \space\rho}W^{\rho}+3W^2W^{\mu}+2W^{\mu}(\nabla_{\rho}W^{\rho})-2W^{\rho}(\nabla^{\mu}W_{\rho})= 0   
\end{eqnarray*}
where
\begin{eqnarray*}
\begin{aligned}
     \mathcal{T}_{\mu \nu}&=2 R_{\mu\nu}W^2-2R_{\nu \rho}\,W^\rho\,W_{\mu}-2R_{\mu \rho}\,W^\rho\,W_{\nu}-4R_{\mu \rho \nu \sigma}\,W^{\rho}\,W^{\sigma}+2\,R\, W_{\mu}W_{\nu}\\
     &-12\,W^2W_{\mu}W_{\nu}-8\,W_{\mu}\,W_{\nu}\nabla_{\rho}W^{\rho}-2(\square W_{\mu})\,W_{\nu}-2\,W_{\mu}\,\square W_{\nu}-4 \nabla^{\rho}W_{\mu}\nabla_{\rho}W_{\nu}\\
     &+ 8(\nabla_\mu W_\rho) W_\nu W^{\rho}+2(\nabla_{\mu}W_{\rho})(\nabla^{\rho}W_\nu)+2(\nabla_{\mu}W_{\nu})(\nabla_{\rho}W^\rho)+2(\nabla_\mu \nabla_\rho W^\rho) W_\nu\\
     &+2 W^\rho(\nabla_\mu \nabla_\rho W_\nu)+8W^\rho W_{\mu} (\nabla_{\nu}W_{\rho})+2(\nabla^{\rho}W_{\mu})(\nabla_\nu W_\rho)-4(\nabla_{\mu}W^{\rho})(\nabla_{\nu}W_\rho)\\
     &+2(\nabla_\nu W_{\mu})(\nabla^{\rho}W_{\rho})+2 W_{\mu}(\nabla_{\nu}\nabla_{\rho}W^{\rho})+2 W^{\rho}(\nabla_{\nu}\nabla_{\rho}W_{\mu})-4 W^{\rho}(\nabla_{\mu}\nabla_{\nu}W_{\rho})\\
     &+g_{\mu \nu}\big(4\,R_{\rho \sigma}\, W^{\rho} W^{\sigma}-R\,W^2+3W^4-8W^{\rho}W^{\sigma}(\nabla_{\sigma}W_{\rho})-2(\nabla_{\rho}W^{\rho})^2\\
     &-4W^{\rho}(\nabla_{\sigma}\nabla_{\rho}W^{\sigma}-\square W_{\rho})+2\nabla^{\sigma}W^{\rho}(2 \nabla_{\sigma}W_{\rho}- \nabla_{\rho}W_{\sigma})\big).
\end{aligned}
\end{eqnarray*}

Considering a mini-superspace approach for the theory in Eq. \eqref{eq:theory} with the ansatz \eqref{eq:metric} and \eqref{eq:ansatzVF}, we obtain (up to boundary terms and overall constants) the following effective Lagrangian
\begin{equation}
    \begin{aligned}
        L_{\rm eff} =& \frac{4 \alpha  w_0^2 \left(N \left(r (2 r w_1+1) f'+r f \left(2 r w_1'+w_1 (3 r w_1+4)\right)+f-1\right)+2 r^2 f w_1 N'\right)}{f N^2}\\&-2 N \left(r f' \left(2 \alpha  f (2 r w_1+1) w_1^2+1\right)+f \left(\alpha  w_1^2 \left(f \left(r \left(4 r w_1'+w_1 (3 r w_1+8)\right)+2\right)-2\right)+1\right)-1\right)\\&-\frac{6 \alpha  r^2 w_0^4}{f^2 N^3}-8 \alpha  r f^2 w_1^2 (r w_1+1) N'.
    \end{aligned}
    \label{eq:effLag}
\end{equation}
The transformation in Eq. \eqref{eq:symmetry} changes the Lagrangian in Eq. \eqref{eq:effLag} only by a boundary term
\begin{equation}
    \Delta L_{\rm eff} = \frac{\dd}{\dd r} \left[ \frac{8 \alpha  \kappa r \left(N (2 r w_1 f+f-1)+\kappa r^2\right)}{N} \right] \equiv \frac{\dd}{\dd r} \mathcal{B},
\end{equation}
and is therefore a continuous symmetry of the static and spherically symmetric system. From Noether's theorem, there is a conserved quantity associated with this symmetry such that
\begin{equation}
    \frac{\dd}{\dd r} \left[ \frac{\partial L_{\rm eff}}{\partial w_0'} \frac{\partial w_0}{\partial \kappa} + \frac{\partial L_{\rm eff}}{\partial w_1'} \frac{\partial w_1}{\partial \kappa} - \frac{\partial \mathcal{B}}{\partial \kappa} \right]\Bigg\rvert_{\kappa=0} = 0,
\end{equation}
from which we obtain Eq. \eqref{eq:consQuant}.

%%%%%%%%%%%%%%%%%%%%%%%%%%%%%%%%%
\bibliography{references}

\begin{thebibliography}{76}%
\makeatletter
\providecommand \@ifxundefined [1]{%
 \@ifx{#1\undefined}
}%
\providecommand \@ifnum [1]{%
 \ifnum #1\expandafter \@firstoftwo
 \else \expandafter \@secondoftwo
 \fi
}%
\providecommand \@ifx [1]{%
 \ifx #1\expandafter \@firstoftwo
 \else \expandafter \@secondoftwo
 \fi
}%
\providecommand \natexlab [1]{#1}%
\providecommand \enquote  [1]{``#1''}%
\providecommand \bibnamefont  [1]{#1}%
\providecommand \bibfnamefont [1]{#1}%
\providecommand \citenamefont [1]{#1}%
\providecommand \href@noop [0]{\@secondoftwo}%
\providecommand \href [0]{\begingroup \@sanitize@url \@href}%
\providecommand \@href[1]{\@@startlink{#1}\@@href}%
\providecommand \@@href[1]{\endgroup#1\@@endlink}%
\providecommand \@sanitize@url [0]{\catcode `\\12\catcode `\$12\catcode `\&12\catcode `\#12\catcode `\^12\catcode `\_12\catcode `\%12\relax}%
\providecommand \@@startlink[1]{}%
\providecommand \@@endlink[0]{}%
\providecommand \url  [0]{\begingroup\@sanitize@url \@url }%
\providecommand \@url [1]{\endgroup\@href {#1}{\urlprefix }}%
\providecommand \urlprefix  [0]{URL }%
\providecommand \Eprint [0]{\href }%
\providecommand \doibase [0]{https://doi.org/}%
\providecommand \selectlanguage [0]{\@gobble}%
\providecommand \bibinfo  [0]{\@secondoftwo}%
\providecommand \bibfield  [0]{\@secondoftwo}%
\providecommand \translation [1]{[#1]}%
\providecommand \BibitemOpen [0]{}%
\providecommand \bibitemStop [0]{}%
\providecommand \bibitemNoStop [0]{.\EOS\space}%
\providecommand \EOS [0]{\spacefactor3000\relax}%
\providecommand \BibitemShut  [1]{\csname bibitem#1\endcsname}%
\let\auto@bib@innerbib\@empty
%</preamble>
\bibitem [{\citenamefont {Lovelock}(1971)}]{Lovelock:1971yv}%
  \BibitemOpen
  \bibfield  {author} {\bibinfo {author} {\bibfnamefont {D.}~\bibnamefont {Lovelock}},\ }\bibfield  {title} {\bibinfo {title} {{The Einstein tensor and its generalizations}},\ }\href {https://doi.org/10.1063/1.1665613} {\bibfield  {journal} {\bibinfo  {journal} {J. Math. Phys.}\ }\textbf {\bibinfo {volume} {12}},\ \bibinfo {pages} {498} (\bibinfo {year} {1971})}\BibitemShut {NoStop}%
\bibitem [{\citenamefont {Padmanabhan}\ and\ \citenamefont {Kothawala}(2013)}]{Padmanabhan:2013xyr}%
  \BibitemOpen
  \bibfield  {author} {\bibinfo {author} {\bibfnamefont {T.}~\bibnamefont {Padmanabhan}}\ and\ \bibinfo {author} {\bibfnamefont {D.}~\bibnamefont {Kothawala}},\ }\bibfield  {title} {\bibinfo {title} {{Lanczos-Lovelock models of gravity}},\ }\href {https://doi.org/10.1016/j.physrep.2013.05.007} {\bibfield  {journal} {\bibinfo  {journal} {Phys. Rept.}\ }\textbf {\bibinfo {volume} {531}},\ \bibinfo {pages} {115} (\bibinfo {year} {2013})},\ \Eprint {https://arxiv.org/abs/1302.2151} {arXiv:1302.2151 [gr-qc]} \BibitemShut {NoStop}%
\bibitem [{\citenamefont {Horndeski}(1974)}]{Horndeski:1974wa}%
  \BibitemOpen
  \bibfield  {author} {\bibinfo {author} {\bibfnamefont {G.~W.}\ \bibnamefont {Horndeski}},\ }\bibfield  {title} {\bibinfo {title} {{Second-order scalar-tensor field equations in a four-dimensional space}},\ }\href {https://doi.org/10.1007/BF01807638} {\bibfield  {journal} {\bibinfo  {journal} {Int. J. Theor. Phys.}\ }\textbf {\bibinfo {volume} {10}},\ \bibinfo {pages} {363} (\bibinfo {year} {1974})}\BibitemShut {NoStop}%
\bibitem [{\citenamefont {Kobayashi}(2019)}]{Kobayashi:2019hrl}%
  \BibitemOpen
  \bibfield  {author} {\bibinfo {author} {\bibfnamefont {T.}~\bibnamefont {Kobayashi}},\ }\bibfield  {title} {\bibinfo {title} {{Horndeski theory and beyond: a review}},\ }\href {https://doi.org/10.1088/1361-6633/ab2429} {\bibfield  {journal} {\bibinfo  {journal} {Rept. Prog. Phys.}\ }\textbf {\bibinfo {volume} {82}},\ \bibinfo {pages} {086901} (\bibinfo {year} {2019})},\ \Eprint {https://arxiv.org/abs/1901.07183} {arXiv:1901.07183 [gr-qc]} \BibitemShut {NoStop}%
\bibitem [{\citenamefont {Ferrara}\ \emph {et~al.}(1996)\citenamefont {Ferrara}, \citenamefont {Khuri},\ and\ \citenamefont {Minasian}}]{Ferrara:1996hh}%
  \BibitemOpen
  \bibfield  {author} {\bibinfo {author} {\bibfnamefont {S.}~\bibnamefont {Ferrara}}, \bibinfo {author} {\bibfnamefont {R.~R.}\ \bibnamefont {Khuri}},\ and\ \bibinfo {author} {\bibfnamefont {R.}~\bibnamefont {Minasian}},\ }\bibfield  {title} {\bibinfo {title} {{M theory on a Calabi-Yau manifold}},\ }\href {https://doi.org/10.1016/0370-2693(96)00270-5} {\bibfield  {journal} {\bibinfo  {journal} {Phys. Lett. B}\ }\textbf {\bibinfo {volume} {375}},\ \bibinfo {pages} {81} (\bibinfo {year} {1996})},\ \Eprint {https://arxiv.org/abs/hep-th/9602102} {arXiv:hep-th/9602102} \BibitemShut {NoStop}%
\bibitem [{\citenamefont {Antoniadis}\ \emph {et~al.}(1997)\citenamefont {Antoniadis}, \citenamefont {Ferrara}, \citenamefont {Minasian},\ and\ \citenamefont {Narain}}]{Antoniadis:1997eg}%
  \BibitemOpen
  \bibfield  {author} {\bibinfo {author} {\bibfnamefont {I.}~\bibnamefont {Antoniadis}}, \bibinfo {author} {\bibfnamefont {S.}~\bibnamefont {Ferrara}}, \bibinfo {author} {\bibfnamefont {R.}~\bibnamefont {Minasian}},\ and\ \bibinfo {author} {\bibfnamefont {K.~S.}\ \bibnamefont {Narain}},\ }\bibfield  {title} {\bibinfo {title} {{R**4 couplings in M and type II theories on Calabi-Yau spaces}},\ }\href {https://doi.org/10.1016/S0550-3213(97)00572-5} {\bibfield  {journal} {\bibinfo  {journal} {Nucl. Phys. B}\ }\textbf {\bibinfo {volume} {507}},\ \bibinfo {pages} {571} (\bibinfo {year} {1997})},\ \Eprint {https://arxiv.org/abs/hep-th/9707013} {arXiv:hep-th/9707013} \BibitemShut {NoStop}%
\bibitem [{\citenamefont {Zwiebach}(1985)}]{Zwiebach:1985uq}%
  \BibitemOpen
  \bibfield  {author} {\bibinfo {author} {\bibfnamefont {B.}~\bibnamefont {Zwiebach}},\ }\bibfield  {title} {\bibinfo {title} {{Curvature Squared Terms and String Theories}},\ }\href {https://doi.org/10.1016/0370-2693(85)91616-8} {\bibfield  {journal} {\bibinfo  {journal} {Phys. Lett. B}\ }\textbf {\bibinfo {volume} {156}},\ \bibinfo {pages} {315} (\bibinfo {year} {1985})}\BibitemShut {NoStop}%
\bibitem [{\citenamefont {Nepomechie}(1985)}]{Nepomechie:1985us}%
  \BibitemOpen
  \bibfield  {author} {\bibinfo {author} {\bibfnamefont {R.~I.}\ \bibnamefont {Nepomechie}},\ }\bibfield  {title} {\bibinfo {title} {{On the Low-energy Limit of Strings}},\ }\href {https://doi.org/10.1103/PhysRevD.32.3201} {\bibfield  {journal} {\bibinfo  {journal} {Phys. Rev. D}\ }\textbf {\bibinfo {volume} {32}},\ \bibinfo {pages} {3201} (\bibinfo {year} {1985})}\BibitemShut {NoStop}%
\bibitem [{\citenamefont {Callan}\ \emph {et~al.}(1986)\citenamefont {Callan}, \citenamefont {Klebanov},\ and\ \citenamefont {Perry}}]{Callan:1986jb}%
  \BibitemOpen
  \bibfield  {author} {\bibinfo {author} {\bibfnamefont {C.~G.}\ \bibnamefont {Callan}, \bibfnamefont {Jr.}}, \bibinfo {author} {\bibfnamefont {I.~R.}\ \bibnamefont {Klebanov}},\ and\ \bibinfo {author} {\bibfnamefont {M.~J.}\ \bibnamefont {Perry}},\ }\bibfield  {title} {\bibinfo {title} {{String Theory Effective Actions}},\ }\href {https://doi.org/10.1016/0550-3213(86)90107-0} {\bibfield  {journal} {\bibinfo  {journal} {Nucl. Phys. B}\ }\textbf {\bibinfo {volume} {278}},\ \bibinfo {pages} {78} (\bibinfo {year} {1986})}\BibitemShut {NoStop}%
\bibitem [{\citenamefont {Candelas}\ \emph {et~al.}(1985)\citenamefont {Candelas}, \citenamefont {Horowitz}, \citenamefont {Strominger},\ and\ \citenamefont {Witten}}]{Candelas:1985en}%
  \BibitemOpen
  \bibfield  {author} {\bibinfo {author} {\bibfnamefont {P.}~\bibnamefont {Candelas}}, \bibinfo {author} {\bibfnamefont {G.~T.}\ \bibnamefont {Horowitz}}, \bibinfo {author} {\bibfnamefont {A.}~\bibnamefont {Strominger}},\ and\ \bibinfo {author} {\bibfnamefont {E.}~\bibnamefont {Witten}},\ }\bibfield  {title} {\bibinfo {title} {{Vacuum configurations for superstrings}},\ }\href {https://doi.org/10.1016/0550-3213(85)90602-9} {\bibfield  {journal} {\bibinfo  {journal} {Nucl. Phys. B}\ }\textbf {\bibinfo {volume} {258}},\ \bibinfo {pages} {46} (\bibinfo {year} {1985})}\BibitemShut {NoStop}%
\bibitem [{\citenamefont {Gross}\ and\ \citenamefont {Sloan}(1987)}]{Gross:1986mw}%
  \BibitemOpen
  \bibfield  {author} {\bibinfo {author} {\bibfnamefont {D.~J.}\ \bibnamefont {Gross}}\ and\ \bibinfo {author} {\bibfnamefont {J.~H.}\ \bibnamefont {Sloan}},\ }\bibfield  {title} {\bibinfo {title} {{The Quartic Effective Action for the Heterotic String}},\ }\href {https://doi.org/10.1016/0550-3213(87)90465-2} {\bibfield  {journal} {\bibinfo  {journal} {Nucl. Phys. B}\ }\textbf {\bibinfo {volume} {291}},\ \bibinfo {pages} {41} (\bibinfo {year} {1987})}\BibitemShut {NoStop}%
\bibitem [{\citenamefont {Charmousis}\ \emph {et~al.}(2012{\natexlab{a}})\citenamefont {Charmousis}, \citenamefont {Gouteraux},\ and\ \citenamefont {Kiritsis}}]{Charmousis:2012dw}%
  \BibitemOpen
  \bibfield  {author} {\bibinfo {author} {\bibfnamefont {C.}~\bibnamefont {Charmousis}}, \bibinfo {author} {\bibfnamefont {B.}~\bibnamefont {Gouteraux}},\ and\ \bibinfo {author} {\bibfnamefont {E.}~\bibnamefont {Kiritsis}},\ }\bibfield  {title} {\bibinfo {title} {{Higher-derivative scalar-vector-tensor theories: black holes, Galileons, singularity cloaking and holography}},\ }\href {https://doi.org/10.1007/JHEP09(2012)011} {\bibfield  {journal} {\bibinfo  {journal} {JHEP}\ }\textbf {\bibinfo {volume} {09}},\ \bibinfo {pages} {011}},\ \Eprint {https://arxiv.org/abs/1206.1499} {arXiv:1206.1499 [hep-th]} \BibitemShut {NoStop}%
\bibitem [{\citenamefont {Glavan}\ and\ \citenamefont {Lin}(2020)}]{glavanEinsteinGaussBonnetGravity4dimensional2020}%
  \BibitemOpen
  \bibfield  {author} {\bibinfo {author} {\bibfnamefont {D.}~\bibnamefont {Glavan}}\ and\ \bibinfo {author} {\bibfnamefont {C.}~\bibnamefont {Lin}},\ }\bibfield  {title} {\bibinfo {title} {Einstein-{{Gauss-Bonnet}} gravity in 4-dimensional space-time},\ }\href {https://doi.org/10.1103/PhysRevLett.124.081301} {\bibfield  {journal} {\bibinfo  {journal} {Physical Review Letters}\ }\textbf {\bibinfo {volume} {124}},\ \bibinfo {pages} {081301} (\bibinfo {year} {2020})},\ \Eprint {https://arxiv.org/abs/1905.03601} {arXiv:1905.03601} \BibitemShut {NoStop}%
\bibitem [{\citenamefont {Gurses}\ \emph {et~al.}(2020)\citenamefont {Gurses}, \citenamefont {Sisman},\ and\ \citenamefont {Tekin}}]{gursesThereNovelEinsteinGaussBonnet2020}%
  \BibitemOpen
  \bibfield  {author} {\bibinfo {author} {\bibfnamefont {M.}~\bibnamefont {Gurses}}, \bibinfo {author} {\bibfnamefont {T.~C.}\ \bibnamefont {Sisman}},\ and\ \bibinfo {author} {\bibfnamefont {B.}~\bibnamefont {Tekin}},\ }\href {https://doi.org/10.1140/epjc/s10052-020-8200-7} {\bibinfo {title} {Is there a novel {{Einstein-Gauss-Bonnet}} theory in four dimensions?}} (\bibinfo {year} {2020}),\ \Eprint {https://arxiv.org/abs/2004.03390} {arXiv:2004.03390} \BibitemShut {NoStop}%
\bibitem [{\citenamefont {Lu}\ and\ \citenamefont {Pang}(2020)}]{luHorndeskiGravityRightarrow42020}%
  \BibitemOpen
  \bibfield  {author} {\bibinfo {author} {\bibfnamefont {H.}~\bibnamefont {Lu}}\ and\ \bibinfo {author} {\bibfnamefont {Y.}~\bibnamefont {Pang}},\ }\bibfield  {title} {\bibinfo {title} {Horndeski {{Gravity}} as \${{D}}{\textbackslash}rightarrow4\$ {{Limit}} of {{Gauss-Bonnet}}},\ }\href {https://doi.org/10.1016/j.physletb.2020.135717} {\bibfield  {journal} {\bibinfo  {journal} {Physics Letters B}\ }\textbf {\bibinfo {volume} {809}},\ \bibinfo {pages} {135717} (\bibinfo {year} {2020})},\ \Eprint {https://arxiv.org/abs/2003.11552} {arXiv:2003.11552} \BibitemShut {NoStop}%
\bibitem [{\citenamefont {Kobayashi}(2020)}]{kobayashiEffectiveScalartensorDescription2020}%
  \BibitemOpen
  \bibfield  {author} {\bibinfo {author} {\bibfnamefont {T.}~\bibnamefont {Kobayashi}},\ }\bibfield  {title} {\bibinfo {title} {Effective scalar-tensor description of regularized {{Lovelock}} gravity in four dimensions},\ }\href {https://doi.org/10.1088/1475-7516/2020/07/013} {\bibfield  {journal} {\bibinfo  {journal} {Journal of Cosmology and Astroparticle Physics}\ }\textbf {\bibinfo {volume} {2020}}\bibfield  {number} {\bibinfo  {number} { (07)},\ \bibinfo {pages} {013}},\ }\Eprint {https://arxiv.org/abs/2003.12771} {arXiv:2003.12771} \BibitemShut {NoStop}%
\bibitem [{\citenamefont {Fernandes}\ \emph {et~al.}(2020)\citenamefont {Fernandes}, \citenamefont {Carrilho}, \citenamefont {Clifton},\ and\ \citenamefont {Mulryne}}]{fernandesDerivationRegularizedField2020}%
  \BibitemOpen
  \bibfield  {author} {\bibinfo {author} {\bibfnamefont {P.~G.~S.}\ \bibnamefont {Fernandes}}, \bibinfo {author} {\bibfnamefont {P.}~\bibnamefont {Carrilho}}, \bibinfo {author} {\bibfnamefont {T.}~\bibnamefont {Clifton}},\ and\ \bibinfo {author} {\bibfnamefont {D.~J.}\ \bibnamefont {Mulryne}},\ }\bibfield  {title} {\bibinfo {title} {Derivation of {{Regularized Field Equations}} for the {{Einstein-Gauss-Bonnet Theory}} in {{Four Dimensions}}},\ }\href {https://doi.org/10.1103/PhysRevD.102.024025} {\bibfield  {journal} {\bibinfo  {journal} {Physical Review D}\ }\textbf {\bibinfo {volume} {102}},\ \bibinfo {pages} {024025} (\bibinfo {year} {2020})},\ \Eprint {https://arxiv.org/abs/2004.08362} {arXiv:2004.08362} \BibitemShut {NoStop}%
\bibitem [{\citenamefont {Hennigar}\ \emph {et~al.}(2020)\citenamefont {Hennigar}, \citenamefont {Kubiznak}, \citenamefont {Mann},\ and\ \citenamefont {Pollack}}]{hennigarTakingLimitGaussBonnet2020}%
  \BibitemOpen
  \bibfield  {author} {\bibinfo {author} {\bibfnamefont {R.~A.}\ \bibnamefont {Hennigar}}, \bibinfo {author} {\bibfnamefont {D.}~\bibnamefont {Kubiznak}}, \bibinfo {author} {\bibfnamefont {R.~B.}\ \bibnamefont {Mann}},\ and\ \bibinfo {author} {\bibfnamefont {C.}~\bibnamefont {Pollack}},\ }\bibfield  {title} {\bibinfo {title} {On {{Taking}} the \${{D}}{\textbackslash}to 4\$ limit of {{Gauss-Bonnet Gravity}}: {{Theory}} and {{Solutions}}},\ }\href {https://doi.org/10.1007/JHEP07(2020)027} {\bibfield  {journal} {\bibinfo  {journal} {Journal of High Energy Physics}\ }\textbf {\bibinfo {volume} {2020}},\ \bibinfo {pages} {27} (\bibinfo {year} {2020})},\ \Eprint {https://arxiv.org/abs/2004.09472} {arXiv:2004.09472} \BibitemShut {NoStop}%
\bibitem [{\citenamefont {Fernandes}(2021)}]{Fernandes:2021dsb}%
  \BibitemOpen
  \bibfield  {author} {\bibinfo {author} {\bibfnamefont {P.~G.~S.}\ \bibnamefont {Fernandes}},\ }\bibfield  {title} {\bibinfo {title} {{Gravity with a generalized conformal scalar field: theory and solutions}},\ }\href {https://doi.org/10.1103/PhysRevD.103.104065} {\bibfield  {journal} {\bibinfo  {journal} {Phys. Rev. D}\ }\textbf {\bibinfo {volume} {103}},\ \bibinfo {pages} {104065} (\bibinfo {year} {2021})},\ \Eprint {https://arxiv.org/abs/2105.04687} {arXiv:2105.04687 [gr-qc]} \BibitemShut {NoStop}%
\bibitem [{\citenamefont {Fernandes}\ \emph {et~al.}(2022)\citenamefont {Fernandes}, \citenamefont {Carrilho}, \citenamefont {Clifton},\ and\ \citenamefont {Mulryne}}]{fernandes4DEinsteinGaussBonnetTheory2022}%
  \BibitemOpen
  \bibfield  {author} {\bibinfo {author} {\bibfnamefont {P.~G.~S.}\ \bibnamefont {Fernandes}}, \bibinfo {author} {\bibfnamefont {P.}~\bibnamefont {Carrilho}}, \bibinfo {author} {\bibfnamefont {T.}~\bibnamefont {Clifton}},\ and\ \bibinfo {author} {\bibfnamefont {D.~J.}\ \bibnamefont {Mulryne}},\ }\bibfield  {title} {\bibinfo {title} {The {{4D Einstein-Gauss-Bonnet Theory}} of {{Gravity}}: {{A Review}}},\ }\href {https://doi.org/10.1088/1361-6382/ac500a} {\bibfield  {journal} {\bibinfo  {journal} {Classical and Quantum Gravity}\ }\textbf {\bibinfo {volume} {39}},\ \bibinfo {pages} {063001} (\bibinfo {year} {2022})},\ \Eprint {https://arxiv.org/abs/2202.13908} {arXiv:2202.13908} \BibitemShut {NoStop}%
\bibitem [{\citenamefont {Carter}(1968)}]{PhysRev.174.1559}%
  \BibitemOpen
  \bibfield  {author} {\bibinfo {author} {\bibfnamefont {B.}~\bibnamefont {Carter}},\ }\bibfield  {title} {\bibinfo {title} {Global structure of the kerr family of gravitational fields},\ }\href {https://doi.org/10.1103/PhysRev.174.1559} {\bibfield  {journal} {\bibinfo  {journal} {Phys. Rev.}\ }\textbf {\bibinfo {volume} {174}},\ \bibinfo {pages} {1559} (\bibinfo {year} {1968})}\BibitemShut {NoStop}%
\bibitem [{\citenamefont {Robinson}(1975)}]{PhysRevLett.34.905}%
  \BibitemOpen
  \bibfield  {author} {\bibinfo {author} {\bibfnamefont {D.~C.}\ \bibnamefont {Robinson}},\ }\bibfield  {title} {\bibinfo {title} {Uniqueness of the kerr black hole},\ }\href {https://doi.org/10.1103/PhysRevLett.34.905} {\bibfield  {journal} {\bibinfo  {journal} {Phys. Rev. Lett.}\ }\textbf {\bibinfo {volume} {34}},\ \bibinfo {pages} {905} (\bibinfo {year} {1975})}\BibitemShut {NoStop}%
\bibitem [{\citenamefont {Mazur}(1982)}]{Mazur:1982db}%
  \BibitemOpen
  \bibfield  {author} {\bibinfo {author} {\bibfnamefont {P.~O.}\ \bibnamefont {Mazur}},\ }\bibfield  {title} {\bibinfo {title} {{PROOF OF UNIQUENESS OF THE KERR-NEWMAN BLACK HOLE SOLUTION}},\ }\href {https://doi.org/10.1088/0305-4470/15/10/021} {\bibfield  {journal} {\bibinfo  {journal} {J. Phys. A}\ }\textbf {\bibinfo {volume} {15}},\ \bibinfo {pages} {3173} (\bibinfo {year} {1982})}\BibitemShut {NoStop}%
\bibitem [{\citenamefont {Chrusciel}\ \emph {et~al.}(2012)\citenamefont {Chrusciel}, \citenamefont {Lopes~Costa},\ and\ \citenamefont {Heusler}}]{Chrusciel:2012jk}%
  \BibitemOpen
  \bibfield  {author} {\bibinfo {author} {\bibfnamefont {P.~T.}\ \bibnamefont {Chrusciel}}, \bibinfo {author} {\bibfnamefont {J.}~\bibnamefont {Lopes~Costa}},\ and\ \bibinfo {author} {\bibfnamefont {M.}~\bibnamefont {Heusler}},\ }\bibfield  {title} {\bibinfo {title} {{Stationary Black Holes: Uniqueness and Beyond}},\ }\href {https://doi.org/10.12942/lrr-2012-7} {\bibfield  {journal} {\bibinfo  {journal} {Living Rev. Rel.}\ }\textbf {\bibinfo {volume} {15}},\ \bibinfo {pages} {7} (\bibinfo {year} {2012})},\ \Eprint {https://arxiv.org/abs/1205.6112} {arXiv:1205.6112 [gr-qc]} \BibitemShut {NoStop}%
\bibitem [{\citenamefont {Eichhorn}\ \emph {et~al.}(2023)\citenamefont {Eichhorn}, \citenamefont {Fernandes}, \citenamefont {Held},\ and\ \citenamefont {Silva}}]{Eichhorn:2023iab}%
  \BibitemOpen
  \bibfield  {author} {\bibinfo {author} {\bibfnamefont {A.}~\bibnamefont {Eichhorn}}, \bibinfo {author} {\bibfnamefont {P.~G.~S.}\ \bibnamefont {Fernandes}}, \bibinfo {author} {\bibfnamefont {A.}~\bibnamefont {Held}},\ and\ \bibinfo {author} {\bibfnamefont {H.~O.}\ \bibnamefont {Silva}},\ }\href@noop {} {\bibinfo {title} {{Breaking black-hole uniqueness at supermassive scales}}} (\bibinfo {year} {2023}),\ \Eprint {https://arxiv.org/abs/2312.11430} {arXiv:2312.11430 [gr-qc]} \BibitemShut {NoStop}%
\bibitem [{\citenamefont {Abbott}\ \emph {et~al.}(2016)\citenamefont {Abbott} \emph {et~al.}}]{LIGOScientific:2016aoc}%
  \BibitemOpen
  \bibfield  {author} {\bibinfo {author} {\bibfnamefont {B.~P.}\ \bibnamefont {Abbott}} \emph {et~al.} (\bibinfo {collaboration} {LIGO Scientific, Virgo}),\ }\bibfield  {title} {\bibinfo {title} {{Observation of Gravitational Waves from a Binary Black Hole Merger}},\ }\href {https://doi.org/10.1103/PhysRevLett.116.061102} {\bibfield  {journal} {\bibinfo  {journal} {Phys. Rev. Lett.}\ }\textbf {\bibinfo {volume} {116}},\ \bibinfo {pages} {061102} (\bibinfo {year} {2016})},\ \Eprint {https://arxiv.org/abs/1602.03837} {arXiv:1602.03837 [gr-qc]} \BibitemShut {NoStop}%
\bibitem [{\citenamefont {Abbott}\ \emph {et~al.}(2017{\natexlab{a}})\citenamefont {Abbott} \emph {et~al.}}]{LIGOScientific:2017vwq}%
  \BibitemOpen
  \bibfield  {author} {\bibinfo {author} {\bibfnamefont {B.~P.}\ \bibnamefont {Abbott}} \emph {et~al.} (\bibinfo {collaboration} {LIGO Scientific, Virgo}),\ }\bibfield  {title} {\bibinfo {title} {{GW170817: Observation of Gravitational Waves from a Binary Neutron Star Inspiral}},\ }\href {https://doi.org/10.1103/PhysRevLett.119.161101} {\bibfield  {journal} {\bibinfo  {journal} {Phys. Rev. Lett.}\ }\textbf {\bibinfo {volume} {119}},\ \bibinfo {pages} {161101} (\bibinfo {year} {2017}{\natexlab{a}})},\ \Eprint {https://arxiv.org/abs/1710.05832} {arXiv:1710.05832 [gr-qc]} \BibitemShut {NoStop}%
\bibitem [{\citenamefont {Abbott}\ \emph {et~al.}(2019)\citenamefont {Abbott} \emph {et~al.}}]{LIGOScientific:2018mvr}%
  \BibitemOpen
  \bibfield  {author} {\bibinfo {author} {\bibfnamefont {B.~P.}\ \bibnamefont {Abbott}} \emph {et~al.} (\bibinfo {collaboration} {LIGO Scientific, Virgo}),\ }\bibfield  {title} {\bibinfo {title} {{GWTC-1: A Gravitational-Wave Transient Catalog of Compact Binary Mergers Observed by LIGO and Virgo during the First and Second Observing Runs}},\ }\href {https://doi.org/10.1103/PhysRevX.9.031040} {\bibfield  {journal} {\bibinfo  {journal} {Phys. Rev. X}\ }\textbf {\bibinfo {volume} {9}},\ \bibinfo {pages} {031040} (\bibinfo {year} {2019})},\ \Eprint {https://arxiv.org/abs/1811.12907} {arXiv:1811.12907 [astro-ph.HE]} \BibitemShut {NoStop}%
\bibitem [{\citenamefont {Amaro-Seoane}\ \emph {et~al.}(2017)\citenamefont {Amaro-Seoane}, \citenamefont {Audley}, \citenamefont {Babak}, \citenamefont {Baker}, \citenamefont {Barausse}, \citenamefont {Bender}, \citenamefont {Berti}, \citenamefont {Binetruy}, \citenamefont {Born}, \citenamefont {Bortoluzzi} \emph {et~al.}}]{amaro2017laser}%
  \BibitemOpen
  \bibfield  {author} {\bibinfo {author} {\bibfnamefont {P.}~\bibnamefont {Amaro-Seoane}}, \bibinfo {author} {\bibfnamefont {H.}~\bibnamefont {Audley}}, \bibinfo {author} {\bibfnamefont {S.}~\bibnamefont {Babak}}, \bibinfo {author} {\bibfnamefont {J.}~\bibnamefont {Baker}}, \bibinfo {author} {\bibfnamefont {E.}~\bibnamefont {Barausse}}, \bibinfo {author} {\bibfnamefont {P.}~\bibnamefont {Bender}}, \bibinfo {author} {\bibfnamefont {E.}~\bibnamefont {Berti}}, \bibinfo {author} {\bibfnamefont {P.}~\bibnamefont {Binetruy}}, \bibinfo {author} {\bibfnamefont {M.}~\bibnamefont {Born}}, \bibinfo {author} {\bibfnamefont {D.}~\bibnamefont {Bortoluzzi}}, \emph {et~al.},\ }\bibfield  {title} {\bibinfo {title} {Laser interferometer space antenna},\ }\href@noop {} {\bibfield  {journal} {\bibinfo  {journal} {arXiv preprint arXiv:1702.00786}\ } (\bibinfo {year} {2017})}\BibitemShut {NoStop}%
\bibitem [{\citenamefont {Barausse}\ \emph {et~al.}(2020)\citenamefont {Barausse} \emph {et~al.}}]{Barausse:2020rsu}%
  \BibitemOpen
  \bibfield  {author} {\bibinfo {author} {\bibfnamefont {E.}~\bibnamefont {Barausse}} \emph {et~al.},\ }\bibfield  {title} {\bibinfo {title} {{Prospects for Fundamental Physics with LISA}},\ }\href {https://doi.org/10.1007/s10714-020-02691-1} {\bibfield  {journal} {\bibinfo  {journal} {Gen. Rel. Grav.}\ }\textbf {\bibinfo {volume} {52}},\ \bibinfo {pages} {81} (\bibinfo {year} {2020})},\ \Eprint {https://arxiv.org/abs/2001.09793} {arXiv:2001.09793 [gr-qc]} \BibitemShut {NoStop}%
\bibitem [{\citenamefont {Ayzenberg}\ \emph {et~al.}(2023)\citenamefont {Ayzenberg} \emph {et~al.}}]{Ayzenberg:2023hfw}%
  \BibitemOpen
  \bibfield  {author} {\bibinfo {author} {\bibfnamefont {D.}~\bibnamefont {Ayzenberg}} \emph {et~al.},\ }\href@noop {} {\bibinfo {title} {{Fundamental Physics Opportunities with the Next-Generation Event Horizon Telescope}}} (\bibinfo {year} {2023}),\ \Eprint {https://arxiv.org/abs/2312.02130} {arXiv:2312.02130 [astro-ph.HE]} \BibitemShut {NoStop}%
\bibitem [{\citenamefont {Akiyama}\ \emph {et~al.}(2019)\citenamefont {Akiyama} \emph {et~al.}}]{EventHorizonTelescope:2019ggy}%
  \BibitemOpen
  \bibfield  {author} {\bibinfo {author} {\bibfnamefont {K.}~\bibnamefont {Akiyama}} \emph {et~al.} (\bibinfo {collaboration} {Event Horizon Telescope}),\ }\bibfield  {title} {\bibinfo {title} {{First M87 Event Horizon Telescope Results. VI. The Shadow and Mass of the Central Black Hole}},\ }\href {https://doi.org/10.3847/2041-8213/ab1141} {\bibfield  {journal} {\bibinfo  {journal} {Astrophys. J. Lett.}\ }\textbf {\bibinfo {volume} {875}},\ \bibinfo {pages} {L6} (\bibinfo {year} {2019})},\ \Eprint {https://arxiv.org/abs/1906.11243} {arXiv:1906.11243 [astro-ph.GA]} \BibitemShut {NoStop}%
\bibitem [{\citenamefont {Psaltis}\ \emph {et~al.}(2020)\citenamefont {Psaltis} \emph {et~al.}}]{EventHorizonTelescope:2020qrl}%
  \BibitemOpen
  \bibfield  {author} {\bibinfo {author} {\bibfnamefont {D.}~\bibnamefont {Psaltis}} \emph {et~al.} (\bibinfo {collaboration} {Event Horizon Telescope}),\ }\bibfield  {title} {\bibinfo {title} {{Gravitational Test Beyond the First Post-Newtonian Order with the Shadow of the M87 Black Hole}},\ }\href {https://doi.org/10.1103/PhysRevLett.125.141104} {\bibfield  {journal} {\bibinfo  {journal} {Phys. Rev. Lett.}\ }\textbf {\bibinfo {volume} {125}},\ \bibinfo {pages} {141104} (\bibinfo {year} {2020})},\ \Eprint {https://arxiv.org/abs/2010.01055} {arXiv:2010.01055 [gr-qc]} \BibitemShut {NoStop}%
\bibitem [{\citenamefont {Kocherlakota}\ \emph {et~al.}(2021)\citenamefont {Kocherlakota} \emph {et~al.}}]{EventHorizonTelescope:2021dqv}%
  \BibitemOpen
  \bibfield  {author} {\bibinfo {author} {\bibfnamefont {P.}~\bibnamefont {Kocherlakota}} \emph {et~al.} (\bibinfo {collaboration} {Event Horizon Telescope}),\ }\bibfield  {title} {\bibinfo {title} {{Constraints on black-hole charges with the 2017 EHT observations of M87*}},\ }\href {https://doi.org/10.1103/PhysRevD.103.104047} {\bibfield  {journal} {\bibinfo  {journal} {Phys. Rev. D}\ }\textbf {\bibinfo {volume} {103}},\ \bibinfo {pages} {104047} (\bibinfo {year} {2021})},\ \Eprint {https://arxiv.org/abs/2105.09343} {arXiv:2105.09343 [gr-qc]} \BibitemShut {NoStop}%
\bibitem [{\citenamefont {Akiyama}\ \emph {et~al.}(2022)\citenamefont {Akiyama} \emph {et~al.}}]{EventHorizonTelescope:2022xqj}%
  \BibitemOpen
  \bibfield  {author} {\bibinfo {author} {\bibfnamefont {K.}~\bibnamefont {Akiyama}} \emph {et~al.} (\bibinfo {collaboration} {Event Horizon Telescope}),\ }\bibfield  {title} {\bibinfo {title} {{First Sagittarius A* Event Horizon Telescope Results. VI. Testing the Black Hole Metric}},\ }\href {https://doi.org/10.3847/2041-8213/ac6756} {\bibfield  {journal} {\bibinfo  {journal} {Astrophys. J. Lett.}\ }\textbf {\bibinfo {volume} {930}},\ \bibinfo {pages} {L17} (\bibinfo {year} {2022})}\BibitemShut {NoStop}%
\bibitem [{\citenamefont {Herdeiro}\ and\ \citenamefont {Radu}(2014)}]{Herdeiro:2014goa}%
  \BibitemOpen
  \bibfield  {author} {\bibinfo {author} {\bibfnamefont {C.~A.~R.}\ \bibnamefont {Herdeiro}}\ and\ \bibinfo {author} {\bibfnamefont {E.}~\bibnamefont {Radu}},\ }\bibfield  {title} {\bibinfo {title} {{Kerr black holes with scalar hair}},\ }\href {https://doi.org/10.1103/PhysRevLett.112.221101} {\bibfield  {journal} {\bibinfo  {journal} {Phys. Rev. Lett.}\ }\textbf {\bibinfo {volume} {112}},\ \bibinfo {pages} {221101} (\bibinfo {year} {2014})},\ \Eprint {https://arxiv.org/abs/1403.2757} {arXiv:1403.2757 [gr-qc]} \BibitemShut {NoStop}%
\bibitem [{\citenamefont {Herdeiro}\ \emph {et~al.}(2016)\citenamefont {Herdeiro}, \citenamefont {Radu},\ and\ \citenamefont {R\'unarsson}}]{Herdeiro:2016tmi}%
  \BibitemOpen
  \bibfield  {author} {\bibinfo {author} {\bibfnamefont {C.}~\bibnamefont {Herdeiro}}, \bibinfo {author} {\bibfnamefont {E.}~\bibnamefont {Radu}},\ and\ \bibinfo {author} {\bibfnamefont {H.}~\bibnamefont {R\'unarsson}},\ }\bibfield  {title} {\bibinfo {title} {{Kerr black holes with Proca hair}},\ }\href {https://doi.org/10.1088/0264-9381/33/15/154001} {\bibfield  {journal} {\bibinfo  {journal} {Class. Quant. Grav.}\ }\textbf {\bibinfo {volume} {33}},\ \bibinfo {pages} {154001} (\bibinfo {year} {2016})},\ \Eprint {https://arxiv.org/abs/1603.02687} {arXiv:1603.02687 [gr-qc]} \BibitemShut {NoStop}%
\bibitem [{\citenamefont {Bakopoulos}\ \emph {et~al.}(2024{\natexlab{a}})\citenamefont {Bakopoulos}, \citenamefont {Charmousis}, \citenamefont {Kanti}, \citenamefont {Lecoeur},\ and\ \citenamefont {Nakas}}]{Bakopoulos:2023fmv}%
  \BibitemOpen
  \bibfield  {author} {\bibinfo {author} {\bibfnamefont {A.}~\bibnamefont {Bakopoulos}}, \bibinfo {author} {\bibfnamefont {C.}~\bibnamefont {Charmousis}}, \bibinfo {author} {\bibfnamefont {P.}~\bibnamefont {Kanti}}, \bibinfo {author} {\bibfnamefont {N.}~\bibnamefont {Lecoeur}},\ and\ \bibinfo {author} {\bibfnamefont {T.}~\bibnamefont {Nakas}},\ }\bibfield  {title} {\bibinfo {title} {{Black holes with primary scalar hair}},\ }\href {https://doi.org/10.1103/PhysRevD.109.024032} {\bibfield  {journal} {\bibinfo  {journal} {Phys. Rev. D}\ }\textbf {\bibinfo {volume} {109}},\ \bibinfo {pages} {024032} (\bibinfo {year} {2024}{\natexlab{a}})},\ \Eprint {https://arxiv.org/abs/2310.11919} {arXiv:2310.11919 [gr-qc]} \BibitemShut {NoStop}%
\bibitem [{\citenamefont {Baake}\ \emph {et~al.}(2024)\citenamefont {Baake}, \citenamefont {Cisterna}, \citenamefont {Hassaine},\ and\ \citenamefont {Hernandez-Vera}}]{Baake:2023zsq}%
  \BibitemOpen
  \bibfield  {author} {\bibinfo {author} {\bibfnamefont {O.}~\bibnamefont {Baake}}, \bibinfo {author} {\bibfnamefont {A.}~\bibnamefont {Cisterna}}, \bibinfo {author} {\bibfnamefont {M.}~\bibnamefont {Hassaine}},\ and\ \bibinfo {author} {\bibfnamefont {U.}~\bibnamefont {Hernandez-Vera}},\ }\bibfield  {title} {\bibinfo {title} {{Endowing black holes with beyond-Horndeski primary hair: An exact solution framework for scalarizing in every dimension}},\ }\href {https://doi.org/10.1103/PhysRevD.109.064024} {\bibfield  {journal} {\bibinfo  {journal} {Phys. Rev. D}\ }\textbf {\bibinfo {volume} {109}},\ \bibinfo {pages} {064024} (\bibinfo {year} {2024})},\ \Eprint {https://arxiv.org/abs/2312.05207} {arXiv:2312.05207 [hep-th]} \BibitemShut {NoStop}%
\bibitem [{\citenamefont {Bakopoulos}\ \emph {et~al.}(2024{\natexlab{b}})\citenamefont {Bakopoulos}, \citenamefont {Chatzifotis},\ and\ \citenamefont {Nakas}}]{Bakopoulos:2023sdm}%
  \BibitemOpen
  \bibfield  {author} {\bibinfo {author} {\bibfnamefont {A.}~\bibnamefont {Bakopoulos}}, \bibinfo {author} {\bibfnamefont {N.}~\bibnamefont {Chatzifotis}},\ and\ \bibinfo {author} {\bibfnamefont {T.}~\bibnamefont {Nakas}},\ }\bibfield  {title} {\bibinfo {title} {{Compact objects with primary hair in shift and parity symmetric beyond Horndeski gravities}},\ }\href {https://doi.org/10.1103/PhysRevD.110.024044} {\bibfield  {journal} {\bibinfo  {journal} {Phys. Rev. D}\ }\textbf {\bibinfo {volume} {110}},\ \bibinfo {pages} {024044} (\bibinfo {year} {2024}{\natexlab{b}})},\ \Eprint {https://arxiv.org/abs/2312.17198} {arXiv:2312.17198 [gr-qc]} \BibitemShut {NoStop}%
\bibitem [{\citenamefont {Charmousis}\ \emph {et~al.}(2025)\citenamefont {Charmousis}, \citenamefont {Iteanu}, \citenamefont {Langlois},\ and\ \citenamefont {Noui}}]{Charmousis:2025xug}%
  \BibitemOpen
  \bibfield  {author} {\bibinfo {author} {\bibfnamefont {C.}~\bibnamefont {Charmousis}}, \bibinfo {author} {\bibfnamefont {S.}~\bibnamefont {Iteanu}}, \bibinfo {author} {\bibfnamefont {D.}~\bibnamefont {Langlois}},\ and\ \bibinfo {author} {\bibfnamefont {K.}~\bibnamefont {Noui}},\ }\href@noop {} {\bibinfo {title} {{Axial perturbations of black holes with primary scalar hair}}} (\bibinfo {year} {2025}),\ \Eprint {https://arxiv.org/abs/2503.22348} {arXiv:2503.22348 [gr-qc]} \BibitemShut {NoStop}%
\bibitem [{\citenamefont {Gervalle}\ and\ \citenamefont {Volkov}(2024)}]{Gervalle:2024yxj}%
  \BibitemOpen
  \bibfield  {author} {\bibinfo {author} {\bibfnamefont {R.}~\bibnamefont {Gervalle}}\ and\ \bibinfo {author} {\bibfnamefont {M.~S.}\ \bibnamefont {Volkov}},\ }\bibfield  {title} {\bibinfo {title} {{Black Holes with Electroweak Hair}},\ }\href {https://doi.org/10.1103/PhysRevLett.133.171402} {\bibfield  {journal} {\bibinfo  {journal} {Phys. Rev. Lett.}\ }\textbf {\bibinfo {volume} {133}},\ \bibinfo {pages} {171402} (\bibinfo {year} {2024})},\ \Eprint {https://arxiv.org/abs/2406.14357} {arXiv:2406.14357 [hep-th]} \BibitemShut {NoStop}%
\bibitem [{\citenamefont {Gervalle}\ and\ \citenamefont {Volkov}(2025)}]{Gervalle:2025awa}%
  \BibitemOpen
  \bibfield  {author} {\bibinfo {author} {\bibfnamefont {R.}~\bibnamefont {Gervalle}}\ and\ \bibinfo {author} {\bibfnamefont {M.~S.}\ \bibnamefont {Volkov}},\ }\href@noop {} {\bibinfo {title} {{Black holes with electroweak hair -- the detailed derivation}}} (\bibinfo {year} {2025}),\ \Eprint {https://arxiv.org/abs/2504.09304} {arXiv:2504.09304 [hep-th]} \BibitemShut {NoStop}%
\bibitem [{\citenamefont {Heisenberg}(2014)}]{Heisenberg:2014rta}%
  \BibitemOpen
  \bibfield  {author} {\bibinfo {author} {\bibfnamefont {L.}~\bibnamefont {Heisenberg}},\ }\bibfield  {title} {\bibinfo {title} {{Generalization of the Proca Action}},\ }\href {https://doi.org/10.1088/1475-7516/2014/05/015} {\bibfield  {journal} {\bibinfo  {journal} {JCAP}\ }\textbf {\bibinfo {volume} {05}},\ \bibinfo {pages} {015}},\ \Eprint {https://arxiv.org/abs/1402.7026} {arXiv:1402.7026 [hep-th]} \BibitemShut {NoStop}%
\bibitem [{\citenamefont {Coll\'eaux}(2020)}]{Colleaux:2020wfv}%
  \BibitemOpen
  \bibfield  {author} {\bibinfo {author} {\bibfnamefont {A.}~\bibnamefont {Coll\'eaux}},\ }\href@noop {} {\bibinfo {title} {{Dimensional aspects of Lovelock-Lanczos gravity}}} (\bibinfo {year} {2020}),\ \Eprint {https://arxiv.org/abs/2010.14174} {arXiv:2010.14174 [gr-qc]} \BibitemShut {NoStop}%
\bibitem [{\citenamefont {Charmousis}\ \emph {et~al.}(2022)\citenamefont {Charmousis}, \citenamefont {Leh\'ebel}, \citenamefont {Smyrniotis},\ and\ \citenamefont {Stergioulas}}]{Charmousis:2021npl}%
  \BibitemOpen
  \bibfield  {author} {\bibinfo {author} {\bibfnamefont {C.}~\bibnamefont {Charmousis}}, \bibinfo {author} {\bibfnamefont {A.}~\bibnamefont {Leh\'ebel}}, \bibinfo {author} {\bibfnamefont {E.}~\bibnamefont {Smyrniotis}},\ and\ \bibinfo {author} {\bibfnamefont {N.}~\bibnamefont {Stergioulas}},\ }\bibfield  {title} {\bibinfo {title} {{Astrophysical constraints on compact objects in 4D Einstein-Gauss-Bonnet gravity}},\ }\href {https://doi.org/10.1088/1475-7516/2022/02/033} {\bibfield  {journal} {\bibinfo  {journal} {JCAP}\ }\textbf {\bibinfo {volume} {02}}\bibfield  {number} {\bibinfo  {number} { (02)},\ \bibinfo {pages} {033}},\ }\Eprint {https://arxiv.org/abs/2109.01149} {arXiv:2109.01149 [gr-qc]} \BibitemShut {NoStop}%
\bibitem [{\citenamefont {Saavedra}\ \emph {et~al.}(2025)\citenamefont {Saavedra}, \citenamefont {Rubilar}, \citenamefont {Fierro}, \citenamefont {Gammon},\ and\ \citenamefont {Mann}}]{Saavedra:2024fzy}%
  \BibitemOpen
  \bibfield  {author} {\bibinfo {author} {\bibfnamefont {A.}~\bibnamefont {Saavedra}}, \bibinfo {author} {\bibfnamefont {G.}~\bibnamefont {Rubilar}}, \bibinfo {author} {\bibfnamefont {O.}~\bibnamefont {Fierro}}, \bibinfo {author} {\bibfnamefont {M.}~\bibnamefont {Gammon}},\ and\ \bibinfo {author} {\bibfnamefont {R.~B.}\ \bibnamefont {Mann}},\ }\bibfield  {title} {\bibinfo {title} {{Neutron stars in 4D Einstein-Gauss-Bonnet gravity}},\ }\href {https://doi.org/10.1103/PhysRevD.111.064071} {\bibfield  {journal} {\bibinfo  {journal} {Phys. Rev. D}\ }\textbf {\bibinfo {volume} {111}},\ \bibinfo {pages} {064071} (\bibinfo {year} {2025})},\ \Eprint {https://arxiv.org/abs/2412.15459} {arXiv:2412.15459 [gr-qc]} \BibitemShut {NoStop}%
\bibitem [{\citenamefont {Bakopoulos}\ \emph {et~al.}(2022)\citenamefont {Bakopoulos}, \citenamefont {Charmousis},\ and\ \citenamefont {Kanti}}]{Bakopoulos:2021liw}%
  \BibitemOpen
  \bibfield  {author} {\bibinfo {author} {\bibfnamefont {A.}~\bibnamefont {Bakopoulos}}, \bibinfo {author} {\bibfnamefont {C.}~\bibnamefont {Charmousis}},\ and\ \bibinfo {author} {\bibfnamefont {P.}~\bibnamefont {Kanti}},\ }\bibfield  {title} {\bibinfo {title} {{Traversable wormholes in beyond Horndeski theories}},\ }\href {https://doi.org/10.1088/1475-7516/2022/05/022} {\bibfield  {journal} {\bibinfo  {journal} {JCAP}\ }\textbf {\bibinfo {volume} {05}}\bibfield  {number} {\bibinfo  {number} { (05)},\ \bibinfo {pages} {022}},\ }\Eprint {https://arxiv.org/abs/2111.09857} {arXiv:2111.09857 [gr-qc]} \BibitemShut {NoStop}%
\bibitem [{\citenamefont {Babichev}\ \emph {et~al.}(2022)\citenamefont {Babichev}, \citenamefont {Charmousis}, \citenamefont {Hassaine},\ and\ \citenamefont {Lecoeur}}]{Babichev:2022awg}%
  \BibitemOpen
  \bibfield  {author} {\bibinfo {author} {\bibfnamefont {E.}~\bibnamefont {Babichev}}, \bibinfo {author} {\bibfnamefont {C.}~\bibnamefont {Charmousis}}, \bibinfo {author} {\bibfnamefont {M.}~\bibnamefont {Hassaine}},\ and\ \bibinfo {author} {\bibfnamefont {N.}~\bibnamefont {Lecoeur}},\ }\bibfield  {title} {\bibinfo {title} {{Conformally coupled theories and their deformed compact objects: From black holes, radiating spacetimes to eternal wormholes}},\ }\href {https://doi.org/10.1103/PhysRevD.106.064039} {\bibfield  {journal} {\bibinfo  {journal} {Phys. Rev. D}\ }\textbf {\bibinfo {volume} {106}},\ \bibinfo {pages} {064039} (\bibinfo {year} {2022})},\ \Eprint {https://arxiv.org/abs/2206.11013} {arXiv:2206.11013 [gr-qc]} \BibitemShut {NoStop}%
\bibitem [{\citenamefont {Fernandes}\ \emph {et~al.}(2021)\citenamefont {Fernandes}, \citenamefont {Carrilho}, \citenamefont {Clifton},\ and\ \citenamefont {Mulryne}}]{Fernandes:2021ysi}%
  \BibitemOpen
  \bibfield  {author} {\bibinfo {author} {\bibfnamefont {P.~G.~S.}\ \bibnamefont {Fernandes}}, \bibinfo {author} {\bibfnamefont {P.}~\bibnamefont {Carrilho}}, \bibinfo {author} {\bibfnamefont {T.}~\bibnamefont {Clifton}},\ and\ \bibinfo {author} {\bibfnamefont {D.~J.}\ \bibnamefont {Mulryne}},\ }\bibfield  {title} {\bibinfo {title} {{Black holes in the scalar-tensor formulation of 4D Einstein-Gauss-Bonnet gravity: Uniqueness of solutions, and a new candidate for dark matter}},\ }\href {https://doi.org/10.1103/PhysRevD.104.044029} {\bibfield  {journal} {\bibinfo  {journal} {Phys. Rev. D}\ }\textbf {\bibinfo {volume} {104}},\ \bibinfo {pages} {044029} (\bibinfo {year} {2021})},\ \Eprint {https://arxiv.org/abs/2107.00046} {arXiv:2107.00046 [gr-qc]} \BibitemShut {NoStop}%
\bibitem [{\citenamefont {Weyl}(1918)}]{Weyl:1918ib}%
  \BibitemOpen
  \bibfield  {author} {\bibinfo {author} {\bibfnamefont {H.}~\bibnamefont {Weyl}},\ }\bibfield  {title} {\bibinfo {title} {{Gravitation and electricity}},\ }\href@noop {} {\bibfield  {journal} {\bibinfo  {journal} {Sitzungsber. Preuss. Akad. Wiss. Berlin (Math. Phys. )}\ }\textbf {\bibinfo {volume} {1918}},\ \bibinfo {pages} {465} (\bibinfo {year} {1918})}\BibitemShut {NoStop}%
\bibitem [{\citenamefont {Jimenez}\ and\ \citenamefont {Koivisto}(2014)}]{jimenezExtendedGaussBonnetGravities2014}%
  \BibitemOpen
  \bibfield  {author} {\bibinfo {author} {\bibfnamefont {J.~B.}\ \bibnamefont {Jimenez}}\ and\ \bibinfo {author} {\bibfnamefont {T.~S.}\ \bibnamefont {Koivisto}},\ }\bibfield  {title} {\bibinfo {title} {Extended {{Gauss-Bonnet}} gravities in {{Weyl}} geometry},\ }\href {https://doi.org/10.1088/0264-9381/31/13/135002} {\bibfield  {journal} {\bibinfo  {journal} {Classical and Quantum Gravity}\ }\textbf {\bibinfo {volume} {31}},\ \bibinfo {pages} {135002} (\bibinfo {year} {2014})},\ \Eprint {https://arxiv.org/abs/1402.1846} {arXiv:1402.1846} \BibitemShut {NoStop}%
\bibitem [{\citenamefont {Bahamonde}\ and\ \citenamefont {Ba{\~n}ados}(2025)}]{bahamondeExactFiveDimensional2025}%
  \BibitemOpen
  \bibfield  {author} {\bibinfo {author} {\bibfnamefont {S.}~\bibnamefont {Bahamonde}}\ and\ \bibinfo {author} {\bibfnamefont {M.}~\bibnamefont {Ba{\~n}ados}},\ }\href {https://doi.org/10.48550/arXiv.2504.02230} {\bibinfo {title} {An exact five dimensional {{Weyl-Geometry Gauss-Bonnet Black Hole}}}} (\bibinfo {year} {2025}),\ \Eprint {https://arxiv.org/abs/2504.02230} {arXiv:2504.02230} \BibitemShut {NoStop}%
\bibitem [{\citenamefont {Barcel{\'o}}\ \emph {et~al.}(2017)\citenamefont {Barcel{\'o}}, \citenamefont {{Carballo-Rubio}},\ and\ \citenamefont {Garay}}]{barceloWeylRelativityNovel2017}%
  \BibitemOpen
  \bibfield  {author} {\bibinfo {author} {\bibfnamefont {C.}~\bibnamefont {Barcel{\'o}}}, \bibinfo {author} {\bibfnamefont {R.}~\bibnamefont {{Carballo-Rubio}}},\ and\ \bibinfo {author} {\bibfnamefont {L.~J.}\ \bibnamefont {Garay}},\ }\bibfield  {title} {\bibinfo {title} {Weyl relativity: {{A}} novel approach to {{Weyl}}'s ideas},\ }\href {https://doi.org/10.1088/1475-7516/2017/06/014} {\bibfield  {journal} {\bibinfo  {journal} {Journal of Cosmology and Astroparticle Physics}\ }\textbf {\bibinfo {volume} {2017}}\bibfield  {number} {\bibinfo  {number} { (06)},\ \bibinfo {pages} {014}},\ }\Eprint {https://arxiv.org/abs/1703.06355} {arXiv:1703.06355} \BibitemShut {NoStop}%
\bibitem [{\citenamefont {Heisenberg}(2017)}]{Heisenberg:2017mzp}%
  \BibitemOpen
  \bibfield  {author} {\bibinfo {author} {\bibfnamefont {L.}~\bibnamefont {Heisenberg}},\ }\bibfield  {title} {\bibinfo {title} {{Generalised Proca Theories}},\ }in\ \href@noop {} {\emph {\bibinfo {booktitle} {{52nd Rencontres de Moriond on Gravitation}}}}\ (\bibinfo {year} {2017})\ pp.\ \bibinfo {pages} {233--241},\ \Eprint {https://arxiv.org/abs/1705.05387} {arXiv:1705.05387 [hep-th]} \BibitemShut {NoStop}%
\bibitem [{\citenamefont {Babichev}\ \emph {et~al.}(2024)\citenamefont {Babichev}, \citenamefont {Charmousis}, \citenamefont {Hassaine},\ and\ \citenamefont {Lecoeur}}]{Babichev:2024krm}%
  \BibitemOpen
  \bibfield  {author} {\bibinfo {author} {\bibfnamefont {E.}~\bibnamefont {Babichev}}, \bibinfo {author} {\bibfnamefont {C.}~\bibnamefont {Charmousis}}, \bibinfo {author} {\bibfnamefont {M.}~\bibnamefont {Hassaine}},\ and\ \bibinfo {author} {\bibfnamefont {N.}~\bibnamefont {Lecoeur}},\ }\bibfield  {title} {\bibinfo {title} {{Global conformal symmetry in scalar-tensor theories}},\ }\href {https://doi.org/10.1142/S0217751X24450039} {\bibfield  {journal} {\bibinfo  {journal} {Int. J. Mod. Phys. A}\ }\textbf {\bibinfo {volume} {39}},\ \bibinfo {pages} {2445003} (\bibinfo {year} {2024})},\ \Eprint {https://arxiv.org/abs/2407.11157} {arXiv:2407.11157 [hep-th]} \BibitemShut {NoStop}%
\bibitem [{\citenamefont {Dom\`enech}\ \emph {et~al.}(2018)\citenamefont {Dom\`enech}, \citenamefont {Mukohyama}, \citenamefont {Namba},\ and\ \citenamefont {Papadopoulos}}]{Domenech:2018vqj}%
  \BibitemOpen
  \bibfield  {author} {\bibinfo {author} {\bibfnamefont {G.}~\bibnamefont {Dom\`enech}}, \bibinfo {author} {\bibfnamefont {S.}~\bibnamefont {Mukohyama}}, \bibinfo {author} {\bibfnamefont {R.}~\bibnamefont {Namba}},\ and\ \bibinfo {author} {\bibfnamefont {V.}~\bibnamefont {Papadopoulos}},\ }\bibfield  {title} {\bibinfo {title} {{Vector disformal transformation of generalized Proca theory}},\ }\href {https://doi.org/10.1103/PhysRevD.98.064037} {\bibfield  {journal} {\bibinfo  {journal} {Phys. Rev. D}\ }\textbf {\bibinfo {volume} {98}},\ \bibinfo {pages} {064037} (\bibinfo {year} {2018})},\ \Eprint {https://arxiv.org/abs/1807.06048} {arXiv:1807.06048 [gr-qc]} \BibitemShut {NoStop}%
\bibitem [{\citenamefont {Babichev}\ \emph {et~al.}(2018)\citenamefont {Babichev}, \citenamefont {Charmousis}, \citenamefont {Esposito-Far\`ese},\ and\ \citenamefont {Leh\'ebel}}]{Babichev:2017lmw}%
  \BibitemOpen
  \bibfield  {author} {\bibinfo {author} {\bibfnamefont {E.}~\bibnamefont {Babichev}}, \bibinfo {author} {\bibfnamefont {C.}~\bibnamefont {Charmousis}}, \bibinfo {author} {\bibfnamefont {G.}~\bibnamefont {Esposito-Far\`ese}},\ and\ \bibinfo {author} {\bibfnamefont {A.}~\bibnamefont {Leh\'ebel}},\ }\bibfield  {title} {\bibinfo {title} {{Stability of Black Holes and the Speed of Gravitational Waves within Self-Tuning Cosmological Models}},\ }\href {https://doi.org/10.1103/PhysRevLett.120.241101} {\bibfield  {journal} {\bibinfo  {journal} {Phys. Rev. Lett.}\ }\textbf {\bibinfo {volume} {120}},\ \bibinfo {pages} {241101} (\bibinfo {year} {2018})},\ \Eprint {https://arxiv.org/abs/1712.04398} {arXiv:1712.04398 [gr-qc]} \BibitemShut {NoStop}%
\bibitem [{\citenamefont {Ben~Achour}\ \emph {et~al.}(2020)\citenamefont {Ben~Achour}, \citenamefont {Liu},\ and\ \citenamefont {Mukohyama}}]{BenAchour:2020wiw}%
  \BibitemOpen
  \bibfield  {author} {\bibinfo {author} {\bibfnamefont {J.}~\bibnamefont {Ben~Achour}}, \bibinfo {author} {\bibfnamefont {H.}~\bibnamefont {Liu}},\ and\ \bibinfo {author} {\bibfnamefont {S.}~\bibnamefont {Mukohyama}},\ }\bibfield  {title} {\bibinfo {title} {{Hairy black holes in DHOST theories: Exploring disformal transformation as a solution-generating method}},\ }\href {https://doi.org/10.1088/1475-7516/2020/02/023} {\bibfield  {journal} {\bibinfo  {journal} {JCAP}\ }\textbf {\bibinfo {volume} {02}},\ \bibinfo {pages} {023}},\ \Eprint {https://arxiv.org/abs/1910.11017} {arXiv:1910.11017 [gr-qc]} \BibitemShut {NoStop}%
\bibitem [{\citenamefont {Abbott}\ \emph {et~al.}(2017{\natexlab{b}})\citenamefont {Abbott} \emph {et~al.}}]{LIGOScientific:2017zic}%
  \BibitemOpen
  \bibfield  {author} {\bibinfo {author} {\bibfnamefont {B.~P.}\ \bibnamefont {Abbott}} \emph {et~al.} (\bibinfo {collaboration} {LIGO Scientific, Virgo, Fermi-GBM, INTEGRAL}),\ }\bibfield  {title} {\bibinfo {title} {{Gravitational Waves and Gamma-rays from a Binary Neutron Star Merger: GW170817 and GRB 170817A}},\ }\href {https://doi.org/10.3847/2041-8213/aa920c} {\bibfield  {journal} {\bibinfo  {journal} {Astrophys. J. Lett.}\ }\textbf {\bibinfo {volume} {848}},\ \bibinfo {pages} {L13} (\bibinfo {year} {2017}{\natexlab{b}})},\ \Eprint {https://arxiv.org/abs/1710.05834} {arXiv:1710.05834 [astro-ph.HE]} \BibitemShut {NoStop}%
\bibitem [{\citenamefont {Dong}\ \emph {et~al.}(2024)\citenamefont {Dong}, \citenamefont {Liu},\ and\ \citenamefont {Liu}}]{Dong:2023xyb}%
  \BibitemOpen
  \bibfield  {author} {\bibinfo {author} {\bibfnamefont {Y.-Q.}\ \bibnamefont {Dong}}, \bibinfo {author} {\bibfnamefont {Y.-Q.}\ \bibnamefont {Liu}},\ and\ \bibinfo {author} {\bibfnamefont {Y.-X.}\ \bibnamefont {Liu}},\ }\bibfield  {title} {\bibinfo {title} {{Polarization modes of gravitational waves in generalized Proca theory}},\ }\href {https://doi.org/10.1103/PhysRevD.109.024014} {\bibfield  {journal} {\bibinfo  {journal} {Phys. Rev. D}\ }\textbf {\bibinfo {volume} {109}},\ \bibinfo {pages} {024014} (\bibinfo {year} {2024})},\ \Eprint {https://arxiv.org/abs/2305.12516} {arXiv:2305.12516 [gr-qc]} \BibitemShut {NoStop}%
\bibitem [{\citenamefont {Barton}\ \emph {et~al.}(2021)\citenamefont {Barton}, \citenamefont {Hartmann}, \citenamefont {Kleihaus},\ and\ \citenamefont {Kunz}}]{Barton:2021wfj}%
  \BibitemOpen
  \bibfield  {author} {\bibinfo {author} {\bibfnamefont {S.}~\bibnamefont {Barton}}, \bibinfo {author} {\bibfnamefont {B.}~\bibnamefont {Hartmann}}, \bibinfo {author} {\bibfnamefont {B.}~\bibnamefont {Kleihaus}},\ and\ \bibinfo {author} {\bibfnamefont {J.}~\bibnamefont {Kunz}},\ }\bibfield  {title} {\bibinfo {title} {{Spontaneously vectorized Einstein-Gauss-Bonnet black holes}},\ }\href {https://doi.org/10.1016/j.physletb.2021.136336} {\bibfield  {journal} {\bibinfo  {journal} {Phys. Lett. B}\ }\textbf {\bibinfo {volume} {817}},\ \bibinfo {pages} {136336} (\bibinfo {year} {2021})},\ \Eprint {https://arxiv.org/abs/2103.01651} {arXiv:2103.01651 [gr-qc]} \BibitemShut {NoStop}%
\bibitem [{\citenamefont {Barton}\ \emph {et~al.}(2022)\citenamefont {Barton}, \citenamefont {Kiefer}, \citenamefont {Kleihaus},\ and\ \citenamefont {Kunz}}]{Barton:2022rkj}%
  \BibitemOpen
  \bibfield  {author} {\bibinfo {author} {\bibfnamefont {S.}~\bibnamefont {Barton}}, \bibinfo {author} {\bibfnamefont {C.}~\bibnamefont {Kiefer}}, \bibinfo {author} {\bibfnamefont {B.}~\bibnamefont {Kleihaus}},\ and\ \bibinfo {author} {\bibfnamefont {J.}~\bibnamefont {Kunz}},\ }\bibfield  {title} {\bibinfo {title} {{Symmetric wormholes in Einstein-vector\textendash{}Gauss\textendash{}Bonnet theory}},\ }\href {https://doi.org/10.1140/epjc/s10052-022-10761-8} {\bibfield  {journal} {\bibinfo  {journal} {Eur. Phys. J. C}\ }\textbf {\bibinfo {volume} {82}},\ \bibinfo {pages} {802} (\bibinfo {year} {2022})},\ \Eprint {https://arxiv.org/abs/2204.08232} {arXiv:2204.08232 [gr-qc]} \BibitemShut {NoStop}%
\bibitem [{\citenamefont {Heisenberg}\ \emph {et~al.}(2017{\natexlab{a}})\citenamefont {Heisenberg}, \citenamefont {Kase}, \citenamefont {Minamitsuji},\ and\ \citenamefont {Tsujikawa}}]{Heisenberg:2017xda}%
  \BibitemOpen
  \bibfield  {author} {\bibinfo {author} {\bibfnamefont {L.}~\bibnamefont {Heisenberg}}, \bibinfo {author} {\bibfnamefont {R.}~\bibnamefont {Kase}}, \bibinfo {author} {\bibfnamefont {M.}~\bibnamefont {Minamitsuji}},\ and\ \bibinfo {author} {\bibfnamefont {S.}~\bibnamefont {Tsujikawa}},\ }\bibfield  {title} {\bibinfo {title} {{Hairy black-hole solutions in generalized Proca theories}},\ }\href {https://doi.org/10.1103/PhysRevD.96.084049} {\bibfield  {journal} {\bibinfo  {journal} {Phys. Rev. D}\ }\textbf {\bibinfo {volume} {96}},\ \bibinfo {pages} {084049} (\bibinfo {year} {2017}{\natexlab{a}})},\ \Eprint {https://arxiv.org/abs/1705.09662} {arXiv:1705.09662 [gr-qc]} \BibitemShut {NoStop}%
\bibitem [{\citenamefont {Heisenberg}\ \emph {et~al.}(2017{\natexlab{b}})\citenamefont {Heisenberg}, \citenamefont {Kase}, \citenamefont {Minamitsuji},\ and\ \citenamefont {Tsujikawa}}]{Heisenberg:2017hwb}%
  \BibitemOpen
  \bibfield  {author} {\bibinfo {author} {\bibfnamefont {L.}~\bibnamefont {Heisenberg}}, \bibinfo {author} {\bibfnamefont {R.}~\bibnamefont {Kase}}, \bibinfo {author} {\bibfnamefont {M.}~\bibnamefont {Minamitsuji}},\ and\ \bibinfo {author} {\bibfnamefont {S.}~\bibnamefont {Tsujikawa}},\ }\bibfield  {title} {\bibinfo {title} {{Black holes in vector-tensor theories}},\ }\href {https://doi.org/10.1088/1475-7516/2017/08/024} {\bibfield  {journal} {\bibinfo  {journal} {JCAP}\ }\textbf {\bibinfo {volume} {08}},\ \bibinfo {pages} {024}},\ \Eprint {https://arxiv.org/abs/1706.05115} {arXiv:1706.05115 [gr-qc]} \BibitemShut {NoStop}%
\bibitem [{\citenamefont {Babichev}\ \emph {et~al.}(2023)\citenamefont {Babichev}, \citenamefont {Charmousis}, \citenamefont {Hassaine},\ and\ \citenamefont {Lecoeur}}]{Babichev:2023dhs}%
  \BibitemOpen
  \bibfield  {author} {\bibinfo {author} {\bibfnamefont {E.}~\bibnamefont {Babichev}}, \bibinfo {author} {\bibfnamefont {C.}~\bibnamefont {Charmousis}}, \bibinfo {author} {\bibfnamefont {M.}~\bibnamefont {Hassaine}},\ and\ \bibinfo {author} {\bibfnamefont {N.}~\bibnamefont {Lecoeur}},\ }\bibfield  {title} {\bibinfo {title} {{Selecting Horndeski theories without apparent symmetries and their black hole solutions}},\ }\href {https://doi.org/10.1103/PhysRevD.108.024019} {\bibfield  {journal} {\bibinfo  {journal} {Phys. Rev. D}\ }\textbf {\bibinfo {volume} {108}},\ \bibinfo {pages} {024019} (\bibinfo {year} {2023})},\ \Eprint {https://arxiv.org/abs/2303.04126} {arXiv:2303.04126 [gr-qc]} \BibitemShut {NoStop}%
\bibitem [{\citenamefont {Charmousis}\ \emph {et~al.}(2012{\natexlab{b}})\citenamefont {Charmousis}, \citenamefont {Copeland}, \citenamefont {Padilla},\ and\ \citenamefont {Saffin}}]{Charmousis:2011bf}%
  \BibitemOpen
  \bibfield  {author} {\bibinfo {author} {\bibfnamefont {C.}~\bibnamefont {Charmousis}}, \bibinfo {author} {\bibfnamefont {E.~J.}\ \bibnamefont {Copeland}}, \bibinfo {author} {\bibfnamefont {A.}~\bibnamefont {Padilla}},\ and\ \bibinfo {author} {\bibfnamefont {P.~M.}\ \bibnamefont {Saffin}},\ }\bibfield  {title} {\bibinfo {title} {{General second order scalar-tensor theory, self tuning, and the Fab Four}},\ }\href {https://doi.org/10.1103/PhysRevLett.108.051101} {\bibfield  {journal} {\bibinfo  {journal} {Phys. Rev. Lett.}\ }\textbf {\bibinfo {volume} {108}},\ \bibinfo {pages} {051101} (\bibinfo {year} {2012}{\natexlab{b}})},\ \Eprint {https://arxiv.org/abs/1106.2000} {arXiv:1106.2000 [hep-th]} \BibitemShut {NoStop}%
\bibitem [{\citenamefont {Charmousis}\ \emph {et~al.}(2012{\natexlab{c}})\citenamefont {Charmousis}, \citenamefont {Copeland}, \citenamefont {Padilla},\ and\ \citenamefont {Saffin}}]{Charmousis:2011ea}%
  \BibitemOpen
  \bibfield  {author} {\bibinfo {author} {\bibfnamefont {C.}~\bibnamefont {Charmousis}}, \bibinfo {author} {\bibfnamefont {E.~J.}\ \bibnamefont {Copeland}}, \bibinfo {author} {\bibfnamefont {A.}~\bibnamefont {Padilla}},\ and\ \bibinfo {author} {\bibfnamefont {P.~M.}\ \bibnamefont {Saffin}},\ }\bibfield  {title} {\bibinfo {title} {{Self-tuning and the derivation of a class of scalar-tensor theories}},\ }\href {https://doi.org/10.1103/PhysRevD.85.104040} {\bibfield  {journal} {\bibinfo  {journal} {Phys. Rev. D}\ }\textbf {\bibinfo {volume} {85}},\ \bibinfo {pages} {104040} (\bibinfo {year} {2012}{\natexlab{c}})},\ \Eprint {https://arxiv.org/abs/1112.4866} {arXiv:1112.4866 [hep-th]} \BibitemShut {NoStop}%
\bibitem [{\citenamefont {Appleby}\ \emph {et~al.}(2012)\citenamefont {Appleby}, \citenamefont {De~Felice},\ and\ \citenamefont {Linder}}]{Appleby:2012rx}%
  \BibitemOpen
  \bibfield  {author} {\bibinfo {author} {\bibfnamefont {S.~A.}\ \bibnamefont {Appleby}}, \bibinfo {author} {\bibfnamefont {A.}~\bibnamefont {De~Felice}},\ and\ \bibinfo {author} {\bibfnamefont {E.~V.}\ \bibnamefont {Linder}},\ }\bibfield  {title} {\bibinfo {title} {{Fab 5: Noncanonical Kinetic Gravity, Self Tuning, and Cosmic Acceleration}},\ }\href {https://doi.org/10.1088/1475-7516/2012/10/060} {\bibfield  {journal} {\bibinfo  {journal} {JCAP}\ }\textbf {\bibinfo {volume} {10}},\ \bibinfo {pages} {060}},\ \Eprint {https://arxiv.org/abs/1208.4163} {arXiv:1208.4163 [astro-ph.CO]} \BibitemShut {NoStop}%
\bibitem [{\citenamefont {Appleby}\ and\ \citenamefont {Linder}(2018)}]{Appleby:2018yci}%
  \BibitemOpen
  \bibfield  {author} {\bibinfo {author} {\bibfnamefont {S.}~\bibnamefont {Appleby}}\ and\ \bibinfo {author} {\bibfnamefont {E.~V.}\ \bibnamefont {Linder}},\ }\bibfield  {title} {\bibinfo {title} {{The Well-Tempered Cosmological Constant}},\ }\href {https://doi.org/10.1088/1475-7516/2018/07/034} {\bibfield  {journal} {\bibinfo  {journal} {JCAP}\ }\textbf {\bibinfo {volume} {07}},\ \bibinfo {pages} {034}},\ \Eprint {https://arxiv.org/abs/1805.00470} {arXiv:1805.00470 [gr-qc]} \BibitemShut {NoStop}%
\bibitem [{\citenamefont {Babichev}\ and\ \citenamefont {Esposito-Farese}(2017)}]{Babichev:2016kdt}%
  \BibitemOpen
  \bibfield  {author} {\bibinfo {author} {\bibfnamefont {E.}~\bibnamefont {Babichev}}\ and\ \bibinfo {author} {\bibfnamefont {G.}~\bibnamefont {Esposito-Farese}},\ }\bibfield  {title} {\bibinfo {title} {{Cosmological self-tuning and local solutions in generalized Horndeski theories}},\ }\href {https://doi.org/10.1103/PhysRevD.95.024020} {\bibfield  {journal} {\bibinfo  {journal} {Phys. Rev. D}\ }\textbf {\bibinfo {volume} {95}},\ \bibinfo {pages} {024020} (\bibinfo {year} {2017})},\ \Eprint {https://arxiv.org/abs/1609.09798} {arXiv:1609.09798 [gr-qc]} \BibitemShut {NoStop}%
\bibitem [{\citenamefont {Deruelle}\ and\ \citenamefont {Farina-Busto}(1990)}]{Deruelle:1989fj}%
  \BibitemOpen
  \bibfield  {author} {\bibinfo {author} {\bibfnamefont {N.}~\bibnamefont {Deruelle}}\ and\ \bibinfo {author} {\bibfnamefont {L.}~\bibnamefont {Farina-Busto}},\ }\bibfield  {title} {\bibinfo {title} {{The Lovelock Gravitational Field Equations in Cosmology}},\ }\href {https://doi.org/10.1103/PhysRevD.41.3696} {\bibfield  {journal} {\bibinfo  {journal} {Phys. Rev. D}\ }\textbf {\bibinfo {volume} {41}},\ \bibinfo {pages} {3696} (\bibinfo {year} {1990})}\BibitemShut {NoStop}%
\bibitem [{\citenamefont {Apostolopoulos}\ \emph {et~al.}(2009)\citenamefont {Apostolopoulos}, \citenamefont {Siopsis},\ and\ \citenamefont {Tetradis}}]{Apostolopoulos:2008ru}%
  \BibitemOpen
  \bibfield  {author} {\bibinfo {author} {\bibfnamefont {P.~S.}\ \bibnamefont {Apostolopoulos}}, \bibinfo {author} {\bibfnamefont {G.}~\bibnamefont {Siopsis}},\ and\ \bibinfo {author} {\bibfnamefont {N.}~\bibnamefont {Tetradis}},\ }\bibfield  {title} {\bibinfo {title} {{Cosmology from an AdS Schwarzschild black hole via holography}},\ }\href {https://doi.org/10.1103/PhysRevLett.102.151301} {\bibfield  {journal} {\bibinfo  {journal} {Phys. Rev. Lett.}\ }\textbf {\bibinfo {volume} {102}},\ \bibinfo {pages} {151301} (\bibinfo {year} {2009})},\ \Eprint {https://arxiv.org/abs/0809.3505} {arXiv:0809.3505 [hep-th]} \BibitemShut {NoStop}%
\bibitem [{\citenamefont {Bilic}(2016)}]{Bilic:2015uol}%
  \BibitemOpen
  \bibfield  {author} {\bibinfo {author} {\bibfnamefont {N.}~\bibnamefont {Bilic}},\ }\bibfield  {title} {\bibinfo {title} {{Randall-Sundrum versus holographic cosmology}},\ }\href {https://doi.org/10.1103/PhysRevD.93.066010} {\bibfield  {journal} {\bibinfo  {journal} {Phys. Rev. D}\ }\textbf {\bibinfo {volume} {93}},\ \bibinfo {pages} {066010} (\bibinfo {year} {2016})},\ \Eprint {https://arxiv.org/abs/1511.07323} {arXiv:1511.07323 [gr-qc]} \BibitemShut {NoStop}%
\bibitem [{\citenamefont {Lidsey}(2013)}]{Lidsey:2009xz}%
  \BibitemOpen
  \bibfield  {author} {\bibinfo {author} {\bibfnamefont {J.~E.}\ \bibnamefont {Lidsey}},\ }\bibfield  {title} {\bibinfo {title} {{Holographic Cosmology from the First Law of Thermodynamics and the Generalized Uncertainty Principle}},\ }\href {https://doi.org/10.1103/PhysRevD.88.103519} {\bibfield  {journal} {\bibinfo  {journal} {Phys. Rev. D}\ }\textbf {\bibinfo {volume} {88}},\ \bibinfo {pages} {103519} (\bibinfo {year} {2013})},\ \Eprint {https://arxiv.org/abs/0911.3286} {arXiv:0911.3286 [hep-th]} \BibitemShut {NoStop}%
\bibitem [{\citenamefont {Cai}\ \emph {et~al.}(2008)\citenamefont {Cai}, \citenamefont {Cao},\ and\ \citenamefont {Hu}}]{Cai:2008ys}%
  \BibitemOpen
  \bibfield  {author} {\bibinfo {author} {\bibfnamefont {R.-G.}\ \bibnamefont {Cai}}, \bibinfo {author} {\bibfnamefont {L.-M.}\ \bibnamefont {Cao}},\ and\ \bibinfo {author} {\bibfnamefont {Y.-P.}\ \bibnamefont {Hu}},\ }\bibfield  {title} {\bibinfo {title} {{Corrected Entropy-Area Relation and Modified Friedmann Equations}},\ }\href {https://doi.org/10.1088/1126-6708/2008/08/090} {\bibfield  {journal} {\bibinfo  {journal} {JHEP}\ }\textbf {\bibinfo {volume} {08}},\ \bibinfo {pages} {090}},\ \Eprint {https://arxiv.org/abs/0807.1232} {arXiv:0807.1232 [hep-th]} \BibitemShut {NoStop}%
\end{thebibliography}%
%%%%%%%%%%%%%%%%%%%%%%%%%%%%

\end{document}